\documentclass[12pt,graphicx]{jpsj3}
\usepackage{amssymb,amsmath}

\def\a{\alpha}
\def\b{\beta}
\def\D{\Delta}

\def\t{\tau}
\def\g{\gamma}
\def\d{\delta}
\def\l{\lambda}

\def\z{\zeta}
\def\ket{\rangle}
\def\bra{\langle}

\def\r{\rho}

\def\s{\sigma}

\def\m{\mu}

\title{Numerical Estimation of the Current Large Deviation Function
in the Asymmetric Simple Exclusion Process with Open Boundary Conditions}
\author{Tetsuya Mitsudo\thanks{E-mail
address:mitsudo@yukawa.kyoto-u.ac.jp} and Shinji Takesue$^1$}
\inst{Yukawa Institute for Theoretical Physics, Kyoto University, Sakyo,
Kyoto 606-8502, Japan \\
Department of Physics, Kyoto University, Sakyo, Kyoto 606-8502, Japan$^1$}

\abst{
We numerically study the large deviation function of the total current, which is
the sum of local currents over all bonds,
for the symmetric and asymmetric simple exclusion processes
with open boundary conditions.
We estimate the generating function by calculating the largest eigenvalue
of the modified transition matrix and by population Monte Carlo simulation.
As a result, we find a number of interesting behaviors not observed in
the exactly solvable cases studied previously as follows.
The even and odd parts of the generating function show different 
system-size dependences.  Different definitions of the current lead to the same
generating function in small systems.  The use of the total current is
important in the Monte Carlo estimation. Moreover, a cusp appears in the 
large deviation function for the asymmetric simple exclusion process.
We also discuss the convergence property of the population Monte Carlo simulation
and find that in a certain parameter region, the convergence is very slow and  
the gap between the largest and second largest eigenvalues of the
modified transition matrix rapidly tends to vanish with the system size.
}

\kword{current large deviation, cumulant generating function,
fluctuation theorem, asymmetric simple exclusion process,
symmetric simple exclusion process, population Monte Carlo,
nonequilibrium, many-body system}

\begin{document}

\maketitle

\section{Introduction}

Current, such as electric current, heat flow, or current of matter,
plays important roles in nonequilibrium systems.
The presence of macroscopic current is a signature of nonequilibrium states.  
Near equilibrium, the current is proportional to the conjugate external
field and the linear coefficient is given by the autocorrelation
function of the current in equilibrium states\cite{Kubo}. 
Beyond the linear regime, although we do not know much about general
properties of the current in strongly interacting systems, current
fluctuations are considered as key quantities to understand the
dynamical behavior of nonequilibrium systems.  
In this respect, the large deviation property of the current is being
studied intensively\cite{Spo,Der07,Tou10} in various systems.
The large deviation function characterizes the probability of the
time-integrated current in the long-time limit and its Legendre
transform gives the cumulant generating function.
Thus, it contains information about not only the mean and the variance
but also higher-order cumulants, which are related to the nonlinear
transport coefficients in the full counting statistics\cite{Sai08}.
In the context of statistical physics, the time-integrated current
called the Helfand moment is related to the Green-Kubo
formula\cite{Hel60}. 
The shear viscosity and the thermal conductivity calculated with the
Helfand moment in the recent simulations are in excellent agreement with
those calculated with the Green-Kubo formula\cite{Vis07-1,Vis07-2}. 
It is also suggested that the current large deviation is relevant to
understanding the dynamical behavior of glasses\cite{Gar07}.

The simple exclusion process (SEP) is probably the simplest and best
studied model of nonequilibrium systems\cite{Schu01,Schad10}.
The SEP is defined on the one-dimensional lattice, where particles are
allowed to move to one of the nearest neighbor sites when the site to
move to is empty.
If the leftward and rightward hopping rates are the same, the model is
called the symmetric simple exclusion process (SSEP).
If the hopping rates are different, the model is called the asymmetric
simple exclusion process (ASEP).
Two extreme cases of the ASEP have special names.
When the difference is as small as the inverse of the system size, the
model is referred to as the weakly ASEP (WASEP).  
The totally ASEP (TASEP) means that one of the hopping rates becomes
zero. 
The ASEP is applied to various phenomena such as traffic
flow\cite{Cho00}, molecular motor transportation\cite{Nis05}, the
process of copying messenger RNA\cite{Mac68}, and sequence
alignment\cite{Bun02}.
The ASEP in the infinite system with a step initial condition can be
mapped to a model of surface growth, and recent experiments show that
the height fluctuations of the growing surface in electroconvection,
which correspond to the current fluctuations in the ASEP, obey the
Tracy-Widom distribution function\cite{Tak10} that appear in the exact
solution of the one-dimensional Karar-Parisi-Zhang
equation\cite{Sas10-2}.

For the ASEP with open boundary conditions, the stationary state is
exactly calculated with the matrix product
method\cite{Der93,Sas98,Uch00} or with the recursion
relation\cite{Schu93}.
In particular, it is clarified that the system has three phases, the
low-density phase, the high-density phase, and the maximum-current
phase, depending on the particle's  entering and exiting rates at the
boundaries.
The transition between the maximum-current phase and the other phases is
second order and that between the low-density and high-density phases is
first order.  Thus, in the latter case, the two phases can coexist on
the phase transition line, where a kink appears between the  low-density
region and the high-density region. The random walk motion of the kink
contributes to the power-law behavior of the power
spectrum\cite{Tak03,Tak08}.

The current large deviation functions for the SSEP with the open
boundary condition were studied by Derrida et al.\cite{Der04,Der07}, who
obtained the scaling form for the generating function.
The TASEP\cite{Der98-1,Der98-2} and the WASEP\cite{Pro09} on a ring, and
the SSEP in the infinite system with the step initial
condition\cite{Der09} were studied by applying the Bethe ansatz.
The direct calculation of the probability of the number of particles
passing through a certain site was performed on the TASEP and the WASEP
in the infinite system  with some special initial
conditions,\cite{Sas08,Sas10} and it turned out that different initial
conditions lead to different types of scaling behavior.
The current moments of the ASEP in the infinitely large system have
recently been studied by Imamura and Sasamoto\cite{Ima11}, who showed
that the calculation of the $n$th moment is reduced to the problem of
the ASEP with a system with particles less than or equal to $n$.
The macroscopic fluctuation theory has been developed as a theory for
the large deviation function\cite{Ber01,Ber06,Bod06,Bod07}.
It is based on the postulate that the large deviation function is
characterized only by the mean and variance of the current.
The macroscopic fluctuation theory is applied to the SSEP\cite{Bod04}
and the WASEP\cite{Bod06}.
In the ASEP, the joint probability of the current and the density is
calculated for some restricted boundary conditions by using the matrix
product method\cite{Dep05}.
For the current through the boundary, the generating function in the
ASEP with open boundary conditions where both input and output are
allowed at each boundary is calculated by using the Bethe
ansatz\cite{Gie11} for the low- and high- density phases, and the
symmetry relation for the current is also discussed\cite{Gie06}.

In addition to the exact calculations, many numerical studies have been
carried out.
A population Monte Carlo method\cite{Iba01} was devised to calculate the
large deviation functions first for discrete time dynamics\cite{Gia06},
and it was extended to continuous time\cite{Lec07}.
By using these methods, the large deviation functions in the zero range
process\cite{Rak08} and in the Kipnis-Marchioro-Pressuti (KMP)
model\cite{Hur10}, which is a simple model of heat conduction, were
computed.
The density matrix renormalization group (DMRG) method was also applied
to obtain the current large deviation for the TASEP with open boundary
conditions\cite{Gor09}.

Despite these extensive studies and the importance of the applications,
the current large deviation function in the ASEP with open boundary
conditions has not yet been fully understood.
Furthermore, the definition of the current varies in the studies.
Some studies consider the current through a bond and other studies
consider the total current, which is the sum of the former over all
bonds.   Although the averages of these two quantities are trivially
equivalent, their fluctuations may behave differently.

In the present study, we numerically calculate the current large
deviation function in the SEP with open boundary conditions and find
several properties that have not been reported in the exactly solvable
cases.
The large deviation function shows a cusp near zero current when the
asymmetry is large.
In both the SSEP and the ASEP, the even part of the generating function
depends linearly on the system size, while the odd part does not.
Our finding indicating that the even-order cumulants are proportional to
system size $L$ differs from that obtained in the preceding
study,\cite{Bod04} where the $n$th-order cumulant is proportional to
$L^{n-1}$. 
We compare the generating functions for the two definitions of the
current and find that they coincide with each other in the direct
evaluation using the largest eigenvalue method, while the population
Monte Carlo method fails when the current through a boundary is employed.
Thus, the use of the total current has an advantage in the Monte Carlo
simulation.
We estimate numerical errors in the Monte Carlo simulation using the
symmetry relation and find that the deviation is usually small.
However, the Monte Carlo simulation sometimes shows an unexpectedly slow
convergence and produces no reliable results.
In those cases, the difference between the largest and second largest
eigenvalues of the modified Master equation decays faster than the
exponential of the system size.

This paper is organized as follows.
In $\S$2, we give a brief review for the SEP and the current large
deviation function, and explain two numerical methods that we use in the
present simulations involving the use of the largest eigenvalue of the
modified Master equation and the population Monte Carlo method.
We also give a short summary of the studies of the current large
deviation for the SSEP and the ASEP under open boundary conditions.
In the next section, we show our numerical results.
In particular, we focus on the difference induced by different
definitions of current.
The convergence problem in the population Monte Carlo method is also discussed.
The last section is devoted to discussion and conclusions.

\section{Model and Algorithm}

\subsection{ASEP and Master equation}

The SEP is a continuous-time
Markov process defined on a one-dimensional chain. 
 We denote $\tau_j=1$ if site $j$ is occupied by a particle and
 $\tau_j=0$ if the site $j$ is empty. 
The configuration of the system of size $L$ is described by the set
$C=\{\tau_1,\tau_2,\dots,\tau_L\}$. 
A particle can hop to the left or right nearest-neighbor site if the
destination site is empty. 
Namely, the particles are subject to the exclusion interaction.
The hopping rate to the right is denoted by  $p_r$ and that to the left
by $p_l$.
If site 1 is empty, a particle enters with rate $\a$, and if the site is 
occupied, the particle exits with rate $\g$.  Similarly, if site $L$ is
empty, a particle enters with rate $\d$, and if the site is occupied,
the particle exits with rate $\b$.

Let us denote the probability for the system to be in configuration $C$
at time $t$ by $P(C,t)$.
By using the transition rate 
from the configuration $C'=\{\t'_1,\dots,\t'_L\}$ to $C=\{\t_1,\dots,\t_L\}$, 
$W_{CC'}$, the time evolution of $P(C,t)$ is described by the Master equation 
\begin{equation}
\frac{{\rm d}}{{\rm d} t}P(C,t)
=\sum_{C'(\ne C)}\left[W_{CC'}P(C',t)-W_{C'C}P(C,t)\right].
\label{master}
\end{equation}
In the ASEP, the transition rate is written as
\begin{eqnarray}
 W_{CC'} &=& w_1 \d_{\t_2\t'_2}\cdots\d_{\t_L\t'_L} \notag \\
 & & +\d_{\t_1\t'_1}\cdots\d_{\t_{j-1}\t'_{j-1}}\sum_{j=1}^{L-1}w_{j,j+1}\d_{\t_{j+2}\t'_{j+2}}\cdots\d_{\t_L\t'_L}
   \notag \\
 & & +\d_{\t_1\t'_1}\cdots\d_{\t_{L-1}\t'_{L-1}}w_L,
\end{eqnarray}
where $\d_{\t_i\t'_i}=(1-\t_i)(1-\t'_i)+\t_i\t'_i$, 
\begin{eqnarray}
w_1=\a(1-\t'_1)\t_1+\g\t'_1(1-\t_1),\\
w_L=\d(1-\t'_L)\t_L+\b\t'_L(1-\t_L),
\end{eqnarray}
 and
\begin{eqnarray}
 \lefteqn{w_{j,j+1}= p_r\t'_j(1-\t'_{j+1})(1-\t_j)\t_{j+1}}\notag\\
& & +p_l(1-\t'_j)\t'_{j+1}\t_j(1-\t_{j+1}).
\end{eqnarray} 
If we denote $r(C)=\sum_{C'(\neq C)}W_{C'C}$, eq. (\ref{master}) is
rewritten as
\begin{equation}
\frac{{\rm d}}{{\rm d} t}P(C,t)
=\sum_{C'(\ne C)}W_{CC'}P(C',t)-r(C)P(C,t).
\label{master2} 
\end{equation}
Thus, $r(C)$ means the rate of transition from $C$ to any other configurations.

The time evolution of the system is decomposed into the set of
one-particle movements that we call a path.
Consider that the system starts with initial configuration $C_0$ at
time $t_0$.
The configuration changes from $C_{i-1}$ to $C_{i}$ when a particle 
changes its position or a particle enters or exits from the system 
at time $t_i$ ($i=1,\dots,K$), 
reaches $C_K$ at $t_{K}$, and remains in 
the same state until time $t$.  Thus, the state at time $t$ is $C=C_K$.
This path is represented as $(C,t;C_K,t_K;\dots;C_1,t_1|C_0,t_0)$. 
Now, we denote the probability that the transition from $C_{i-1}$ to
$C_{i}$ occurs between time $t_i$ and $t_i+{\rm d}t_{i}$ for $i=1,2,\dots,K$ by
\begin{equation}
 P(C,t;C_K,t_K;\dots;C_1,t_1|C_0,t_0){\rm d}t_K\dots {\rm d}t_{2}{\rm d}t_{1}.
\end{equation}
Then the path probability density
$P(C,t;C_{K},t_{K};\dots;C_1,t_1|C_0,t_0)$ for $K\ge 1$ is written as
\begin{eqnarray}
\lefteqn{ P(C,t;C_K,t_K;\dots;C_1,t_1|C_0,t_0)}\notag\\
&=&\delta_{CC_K}e^{-r(C_K)(t-t_K)}W_{C_KC_{K-1}}\notag \\
& & \cdots
 e^{-r(C_1)(t_2-t_1)}W_{C_1C_0}e^{-r(C_0)(t_1-t_0)} \notag\\
&=&\delta_{CC_K}\exp\left[-\sum_{i=0}^{K}r(C_i)(t_{i+1}-t_{i})\right]
 \prod_{j=0}^{K-1}W_{C_{j+1}C_j}, 
\end{eqnarray}
where $t_{K+1}$ should be understood as $t$.
Because $K$ represents the number of one-particle movements, if $K=0$, 
$P(C,t;C_K,t_K;\dots;C_1,t_1|C_0,t_0)$ should be interpreted as representing 
the probability that no
transitions occur at the time interval $[t_0,t]$, which equals
$\delta_{CC_0}e^{-r(C_0)(t-t_0)}$. 
The relationship between the conditional probability $P(C,t|C_0,t_0)$ and
the path probability density is given by
\begin{eqnarray}
\lefteqn{ P(C,t|C_0,t_0)}\notag\\
&=&\sum_{K=0}^{\infty}\sum_{C_1,C_2,\dots,C_K}\int_{t_0}^{t}{\rm d}t_{K}
\int_{t_0}^{t_{K}}{\rm d}t_{K-1}\cdots\int_{t_0}^{t_2}{\rm d}t_1\notag\\
& & P(C,t;C_K,t_K;\dots;C_1,t_1|C_0,t_0).
\end{eqnarray}
If we differentiate the above equation with respect to $t$, we reobtain
the Master equation.

\subsection{Large deviation function and generating function}

Let us denote the probability that the mean current takes the value $q$
as
\begin{equation}
 q=\frac{1}{t}\sum_{i=0}^{K-1} Q(C_{i+1},C_{i})
\end{equation}
by $P(q)$.
The microscopic current $Q(C_{i+1},C_i)$ takes the value $1(-1)$ when the
particle moves forward (backward) during the configuration change
from $C_i$ to $C_{i+1}$.
A precise definition of the microscopic current is given in the next
subsection.
When $t$ is large, this probability asymptotically behaves as
\begin{equation}
 P(q)\simeq e^{-tf(q)},
\end{equation}
where $f(q)$ is called the large deviation function.  
As a result of the law of large number, 
$f(q)$ must have a minimum at a certain value $q^*$ and satisfy $f(q^*)=0$.
The Legendre transform of
$f(q)$ is called the generating function, which we write
\begin{equation}
 \mu(\lambda)=\max_{q}\left[\lambda q-f(q)\right].
\end{equation}
The generating function satisfies $\mu(0)=0$ that corresponds to $f(q^*)=0$.
Similarly, the large deviation function is given by the
Legendre transform of the generating function as
\begin{equation}
  f(q)=\max_{\l}\left[q\l-\m(\l)\right].
\end{equation}
The generating function is also rewritten as
\begin{equation}
\label{braket}
 e^{\m(\l)t}=\bra e^{\l q t} \ket
\end{equation}
for large $t$.
This means that the generating function is expanded as 
\begin{equation}
\label{expand}
 \m(\l)=\sum_{n=1}^{\infty}\frac{1}{n!}\m^{(n)}(0)\l^n=
  \sum_{n=1}^{\infty}\frac{1}{n!}\bra q^n\ket_c t^{n-1}\l^n
\end{equation}
with respect to the cumulants $\bra q^n\ket_c$, or 
the cumulants are generated as
\begin{equation}
 \bra q^n \ket_c=\frac{1}{t^{n-1}} \left. \frac{\partial^n \m(\l)}{\partial
  \l^n}\right|_{\l=0},
\end{equation} 
which is the origin of the name of the function.

The expectation in the rhs of eq. (\ref{braket}) is taken with
respect to the path probability.  
Thus, we obtain
\begin{eqnarray}
\lefteqn{ \bra e^{\l q t} \ket
 = \sum_{K=0}^{\infty} \sum_{C_{K},\cdots,C_1,C}
  \int_{t_0}^{t}{\rm d}t_{K}\int_{t_0}^{t_{K}}{\rm d}
  t_{K-1}\cdots\int_{t_0}^{t_2} {\rm d}t_{1}}\notag \\
& &\d_{C_{K}C}
  e^{\l\sum_{i=0}^{K-1} Q(C_{i+1},C_{i})} e^{-r(C_K)(t-t_K)}
  \nonumber \\
& &
 \prod_{i=0}^{K-1}W_{C_{i+1}C_{i}}e^{-r(C_i)(t_{i+1}-t_{i})}.
\end{eqnarray}
If we introduce $W^{\l}_{CC'}=W_{CC'}e^{\l Q(C,C')}$ and 
\begin{eqnarray}
\lefteqn{P^{\l}(C,t|C_0,t_0)}\nonumber\\
&=&\sum_{K=0}^{\infty} \sum_{C_{K},\cdots,C_1}
  \int_{t_0}^{t}{\rm d}t_{K}\int_{t_0}^{t_{K}}{\rm d}
  t_{K-1}\cdots\int_{t_0}^{t_2} {\rm d}t_{1} \nonumber \\ 
& &\d_{C_{K}C}
  e^{-\sum_{i=1}^{K}r(C_i)(t_{i+1}-t_{i})}
\prod_{j=0}^{K-1}W^{\l}_{C_{j+1}C_{j}},
\end{eqnarray}
the generating function is obtained as
\begin{equation}
 \m(\l)=\lim_{t\to\infty}\frac{1}{t}\log\left[\sum_{C}P^{\l}(C,t|C_0,t_0)\right],
\label{mula}
\end{equation}
or we may assume some initial distribution $P_{0}(C_0)$ and substitute
\begin{equation}
 P^{\l}(C,t)=\sum_{C_0}P^{\l}(C,t|C_0,t_0)P_{0}(C_0)
\end{equation}
in place of $P^{\l}(C,t|C_0,t_0)$ in the rhs of eq. (\ref{mula}).
The time evolution of $P^{\l}(C,t)$ obeys the modified master equation
\begin{eqnarray}
  \frac{{\rm d}}{{\rm d} t}P^{\l}(C,t)=
\sum_{C'\ne C}W^{\l}_{CC'}P^{\l}(C',t)-r(C)P^{\l}(C,t).
\label{masterL}
\end{eqnarray}
We note here that because $\sum_{C'(\ne C)}W^{\l}_{CC'}\ne r(C)$, this
equation cannot be interpreted as an evolution of probability.  Namely,
$\sum_{C}P^{\l}(C,t)$ varies in time.
In the vector form, eq. (\ref{masterL}) is written as
\begin{equation}
 \frac{\rm d}{{\rm d}t}\mathsf{P}^{\l}(t)=\mathsf{W}^{\l}\mathsf{P}(t),
\end{equation}
where 
$\mathsf{P}^{\l}(t)$ is the vector whose $C$th element is $P^{\l}(C,t)$, 
and the $\l$-modified
transition matrix $\mathsf{W}^{\l}$ is defined as
\begin{eqnarray}
 \left(\mathsf{W}^{\l}\right)_{CC'} &=& \left\{
\begin{array}{l} W^{\l}_{CC'} \; \mbox{for} \;C\neq C' \\
-r(C) \; \mbox{for} \; C=C' \end{array}\right. .
\end{eqnarray} 
Because eq. (\ref{masterL}) is linear, the solution is spectrally
decomposed into
\begin{equation}
 P^{\l}(C,t)=\sum_{n}e^{\z_n t}\psi_n(C)\sum_{C'}\psi'_n(C')P_0(C'),
\end{equation}
where $\z_n$ is the eigenvalue and $\psi'_n(C)$ and $\psi_n(C)$ are the 
corresponding left and right eigenvectors of
$\mathsf{W}^{\l}$.
In the long time limit, only the term with the largest eigenvalue
$\z_{\mathrm{max}}$ is dominant, and accordingly the generating function becomes equal
to $\z_{\mathrm{max}}$.
Thus, we can estimate the generating function by numerically calculating
the largest eigenvalue of the $\l$-modified transition matrix if exact 
calculations are not possible.
This method is powerful but restricted to small systems because of
memory limitations.

\subsection{Definition of the microscopic current}

There are two different definitions of current.
One is the current through the boundary of the system.  From the
equation of continuity, this current determines changes of the
amount of the conserved quantity in the system. 
The other is the total current given as the spatial integration
of the current density.
It is natural to use the total current in the linear response theory
when we consider the response to a static and uniform external field
or thermal force.

Both definitions appear in the literature of SEP.
In the studies of current large deviations of the SEP with periodic 
boundaries, the total current is employed by Derrida and
Lebowitz\cite{Der98-1}, Derrida and Appert,\cite{Der98-2} and Prolhac and
Mallick\cite{Pro09}.  
On the contrary, the current through the left boundary is used by Derrida
 et al. \cite{Der04,Der07} or de Gier and Essler\cite{Gie06,Gie11}, who study
the systems with open boundaries.

We let $Q_A$ denote the total current defined as
\begin{eqnarray}
\label{currenta}
Q_A(C_{i+1},C_i) &=& \sum_{j=1}^{L-1} \left[ -\t^i_j(1-\t^i_{j+1})
	       (1-\t^{i+1}_j)\t^{i+1}_{j+1}\right.\notag \\
& &  \mbox{}\left. +(1-\t^i_j)\t^i_{j+1}\t^{i+1}_j(1-\t^{i+1}_{j+1})
	    \right]  \notag \\
& & \mbox{}-(1-\t^i_1)\t^{i+1}_1+\t^i_1(1-\t^{i+1}_1)\notag \\
& & +(1-\t^i_L)\t^{i+1}_L-\t^i_L(1-\t^{i+1}_L),
\end{eqnarray}
and $Q_B$ denote the current through the left boundary as
\begin{eqnarray}
\label{currentb}
 Q_B(C_{i+1},C_i) &=&
-(1-\t^i_1)\t^{i+1}_1+\t^i_1(1-\t^{i+1}_1).
\end{eqnarray}
For each current, we define the mean current $q_X$ ($X=A$ or $B$) as
\begin{equation}
 q_X=\frac{1}{t}\sum_{i=0}^{K-1}Q_X(C_{i+1},C_i), 
\end{equation}
and the generating function $\m_X(\l)$ as
\begin{equation}
 e^{\m_X(\l)}=\bra e^{\l q_X t}\ket.
\end{equation}
We note that the positive direction of the currents
(\ref{currenta}) and (\ref{currentb}) is defined as leftward in this paper.
We use the character $q$ for a current if we do not need to specify
$q_A$ or $q_B$, or for a current in general.
Since the system goes to a stationary state in the long-time limit, we obtain
\begin{equation}
\label{qAB}
 \bra q_A\ket = (L+1) \bra q_B\ket.
\end{equation}
Thus, the two definitions are equivalent with respect to 
expectation values except the trivial factor.

Our question here is whether the large deviation properties of the
two currents appear the same or not.  Theoretically, they are considered
equivalent as follows\cite{referee}.  If a path is fixed, most
particles that enter the system through a boundary exit from either
boundary and the contribution from such particles to $q_A$ is exactly
$L+1$ times the contribution to $q_B$.  The number of such particles 
is proportional to $t$.  On the contrary,
the particles initially present in the system or 
those remaining in the system at time $t$ contribute to $q_A$ and $q_B$
differently.  However, the number of those particles should be $t$-independent
and the contribution from those particles
is negligible for a large time.  Thus, $q_A$ and $q_B$
are supposed to be essentially the same. 
Despite such consideration, how they appear
in the numerical experiment is a different problem.  Actually, as will
be clarified later, the use of the total current has an advantage in the
Monte Carlo simulations.

\subsection{Monte Carlo simulations}

If the system size is so large that the direct evaluation of the largest
eigenvalue is not possible, we can employ a Monte Carlo method.  
However, because $P_{\lambda}(C,t)$ is not a probability distribution function,
conventional Monte Carlo methods cannot be used.
In fact, an algorithm based on the diffusion Monte Carlo method
is devised by Giardin\`{a} et al.\cite{Gia06} and extended to 
a continuous-time case by Lecomte and Tailleur\cite{Lec07}.  
In those algorithms, besides the conventional Monte Carlo dynamics,
we introduce clones of configuration $C$.
The clones 
are duplicated or pruned with the rate
$y(C)=e^{(r_{\l}(C)-r(C))\D t}$, where $\D t$ is the Poisson-distributed
waiting time and its distribution function is given as
\begin{equation}
 r_{\l}(C)e^{-r_{\l}(C)\D t},
\end{equation}
where
\begin{equation}
 r_{\l}(C)=\sum_{C'\neq C}W^{\l}_{CC'}.
\end{equation}
Thus, the mean waiting time is $1/r_{\l}(C)$.  
Since the number of clones must be limited in the computer, we
actually need an additional algorithm to keep the number of clones constant.
If the clone is to be pruned, we choose a random clone and copy over the pruned
one, and if the clone is to be duplicated, we copy the clone on the randomly
chosen clone.

The precise algorithm is as follows. 
\begin{enumerate}
\item Set initial conditions for $N$ clones.
\item Choose which clone to evolve.  Here, we name it $c^{\alpha}$.
First, the clones are chosen orderly.  After all the clones
      evolve in the initial step, the clone with the earliest time is chosen.
\item Calculate the transition probability $W^{\l}_{CC'}/r_{\l}(C)$,
      and let a transition take place. The time interval $\D t$ is
      chosen from the Poisson law. 
\item Calculate $y(C)$ and set $y=[y(C)+\xi]$, where $\xi$ is a uniform random number 
in $0\leq\xi<1$ and $[x]$ is the integer part of $x$.  If $y=1$, the clone
$c^{\a}$ is preserved; if $y=0$, the clone $c^{\a}$ is erased and
overwritten by another randomly chosen clone; and if $y>1$,
$y-1$ clones are chosen randomly and overwritten with the clone $c^{\a}$.
\end{enumerate}

Thus, the total number of clones is kept constant.  
We denote $y$ at the $i$th time step by $y_i$ and define $X_i=(N+y_i-1)/N$.  
Thus, the generating function is given by the following formula in the long-time limit:\cite{Lec07}
\begin{equation}
\label{direct}
 \m(\l)=\frac{1}{t}\ln(X_1\cdots X_K),
\end{equation}
where $K$ is the number of state changes.

Lecomte and Tailleur\cite{Lec07} introduced another method
called thermodynamic integration.
Because the generating function is written as
\begin{equation}
 \m(\l)=\frac{1}{t}\ln\bra e^{q\l t}\ket,
\end{equation}
its derivative is 
\begin{equation}
 \m'(\l)=\frac{\bra qe^{q\l t}\ket}{\bra e^{q\l t}\ket}=\bar{q}^{\l},
\end{equation}
which can be interpreted as the average value of $q$ with respect to 
the population of clones.
Thus, obtaining $\bar{q}^{\l}$ numerically and integrating it from $0$ to $\l$, 
we obtain
\begin{equation}
 \m(\l)=\int_0^{\l}{\rm d}\l' \bar{q}^{\l'}.
\end{equation}
This thermodynamic integration gives a smoother result than the direct
evaluation (\ref{direct}).
In this study, we use both methods and adopt a better result.

\subsection{Known results and the symmetry relations}

There are some known results about the current large deviation for the
SEP with open boundaries.

The cumulants of the current in the SSEP were
calculated \cite{Bod04} under the conjecture of the additivity principle as
\begin{eqnarray}
\label{bod04}
\bra q_b\ket &=& \frac{I_1}{L} \\
\bra q_b^2\ket_c &=& \frac{tI_2}{LI_1} \\
\bra q_b^3\ket_c &=& \frac{t^2}{L}\frac{3(I_3I_1-I_2^2)}{I_1^3} \\
\bra
 q_b^4\ket_c &=& \frac{t^3}{L}\frac{3(5I_4I_1^2-14I_1I_2I_3+9I_2^3)}{I_1^5} \\
I_n &=& \int_{\r_b}^{\r_a}{\rm d}\r D(\r)\s(\r)^{n-1}.
\end{eqnarray}
Here, the macroscopic parameters are $D(\r)=1$ and
$\s(\r)=2\r(1-\r)$, and the boundary parameters are
$\r_a=\frac{\a}{\a+\g}$ and $\r_b=\frac{\d}{\b+\d}$ .
Note that all the cumulants are proportional to $1/L$.

Next, a fluctuation-theorem-like symmetry was found for 
the current large deviation function for the SSEP \cite{Der04,Der07} as 
\begin{equation}
\label{symmetryfq}
 f(q_B)-f(-q_B)=A_{B}q_B,
\end{equation}
where $A_{B}$ is the constant written as
\begin{equation}
\label{abssep}
 A_{B}=\ln{\frac{\a\b}{\g\d}}.
\end{equation}
De Gier and Essler\cite{Gie06} generalized this to the ASEP, where 
the same equation (\ref{symmetryfq}) holds with the coefficient $A_{B}$
given as
\begin{equation}
\label{abasep}
 A_{B}= \ln{\frac{\a\b}{\g\d}}+(L-1)\ln{\frac{p_l}{p_r}}.
\end{equation}
The coefficient $A_{B}$ was derived from the detailed balance condition
given by Enaud and Derrida\cite{Ena04}, which is of the form
\begin{equation}
\label{dbalance}
 \frac{\a\b}{\g\d}\left(\frac{p_l}{p_r}\right)^{L-1}=1.
\end{equation}
In the same manner,
we can further extend the symmetry relation for the mean total current $q_A$ as
\begin{equation}
 f(q_A)-f(-q_A)=A_{A}q_A,
\end{equation}
where
\begin{equation}
\label{aaasep}
 A_{A}=\frac{1}{L+1}A_{B}.
\end{equation}

We briefly review other studies related to our study.
The WASEP has been well studied. 
Bodineau and Derrida \cite{Bod06} assumed the macroscopic
fluctuation theory and extended the results on the WASEP to the strong
asymmetry limit, where they estimated the boundary effect on the local
Gaussian fluctuation.
Depken and Stinchcombe\cite{Dep05} calculated the joint
probability of the density and the current in the TASEP when $\g=\d=0$.
Prolhac and Mallick\cite{Pro09} calculated the cumulants of the current 
in the WASEP with periodic boundary conditions 
using the Bethe ansatz.

Now, we see that the large deviation for the total current
in the SEP with open boundary conditions has not been well studied.

\section{Simulation Results}

\subsection{Open boundary SSEP}

\begin{figure}
\begin{center}
\includegraphics[scale=0.5]{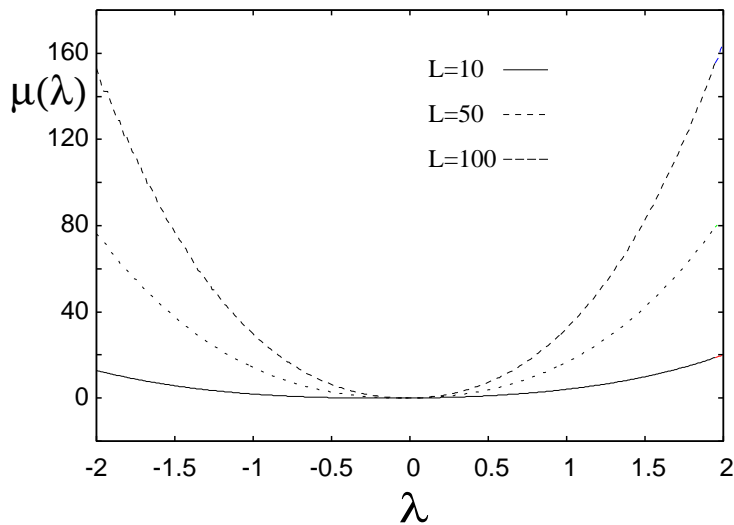}
\caption{Generating function $\m_A(\l)$ calculated by the largest
 eigenvalue method when $L=10$ and those calculated by the Monte Carlo method when
 $L=50$ and $100$. The line is interpolated to bring clearness to the graph.}
\label{fig1}
\end{center}
\end{figure}

First, we show the results on the open boundary SSEP where the hopping rates
are set as $p_l=p_r=1.0$.
Figure \ref{fig1} shows some examples of the generating function of $q_A$ calculated by the two
methods; the largest eigenvalue method ($L=10$) and the Monte Carlo method
($L=50$ and $100$).
The boundary parameters are chosen as $(\a,\b,\g,\d)=(0.1,0.2,0.9,0.8)$
just for illustration.
The generating functions depict parabola-like curves, whose width
decreases as the system size increases.

\begin{figure}
\begin{center}
\includegraphics[scale=0.5]{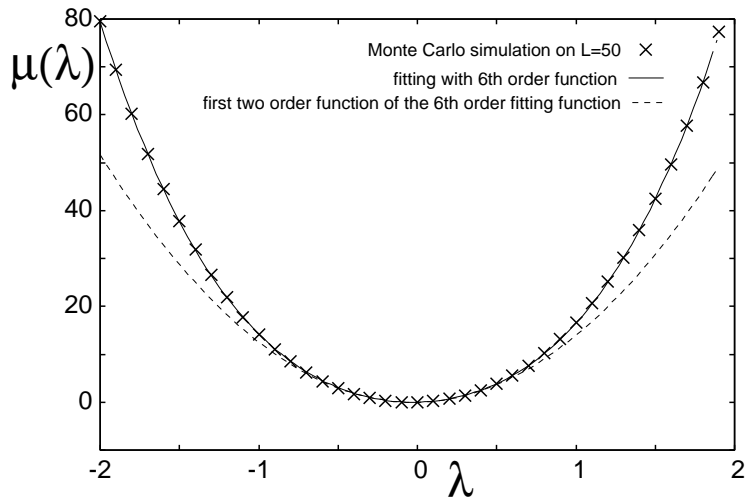}
\caption{Comparison between results obtained by the Monte Carlo simulation with $L=50$
 and fitting function of the 6th-order. The broken line shows the
 quadratic part of the 6th-order fitting function.}
\label{fig2}
\end{center}
\end{figure}

We fit $\m_A(\l)$ to the power series up to the
sixth order using the least-squares method and examine the system size
dependence of the coefficients, which are the cumulants of the current.
The fitting is performed in the range $-2\le\l\le 2$.  The first-order
cumulant is fixed to the steady current.
As seen from Fig. \ref{fig2}, the fitting is successfully performed in the entire range
with $L=50$, whereas the fitting up to the second order significantly
deviates from the simulation results in the range $|\lambda|\gtrsim 0.5$.

\begin{figure}
\begin{center}
\includegraphics[scale=0.35]{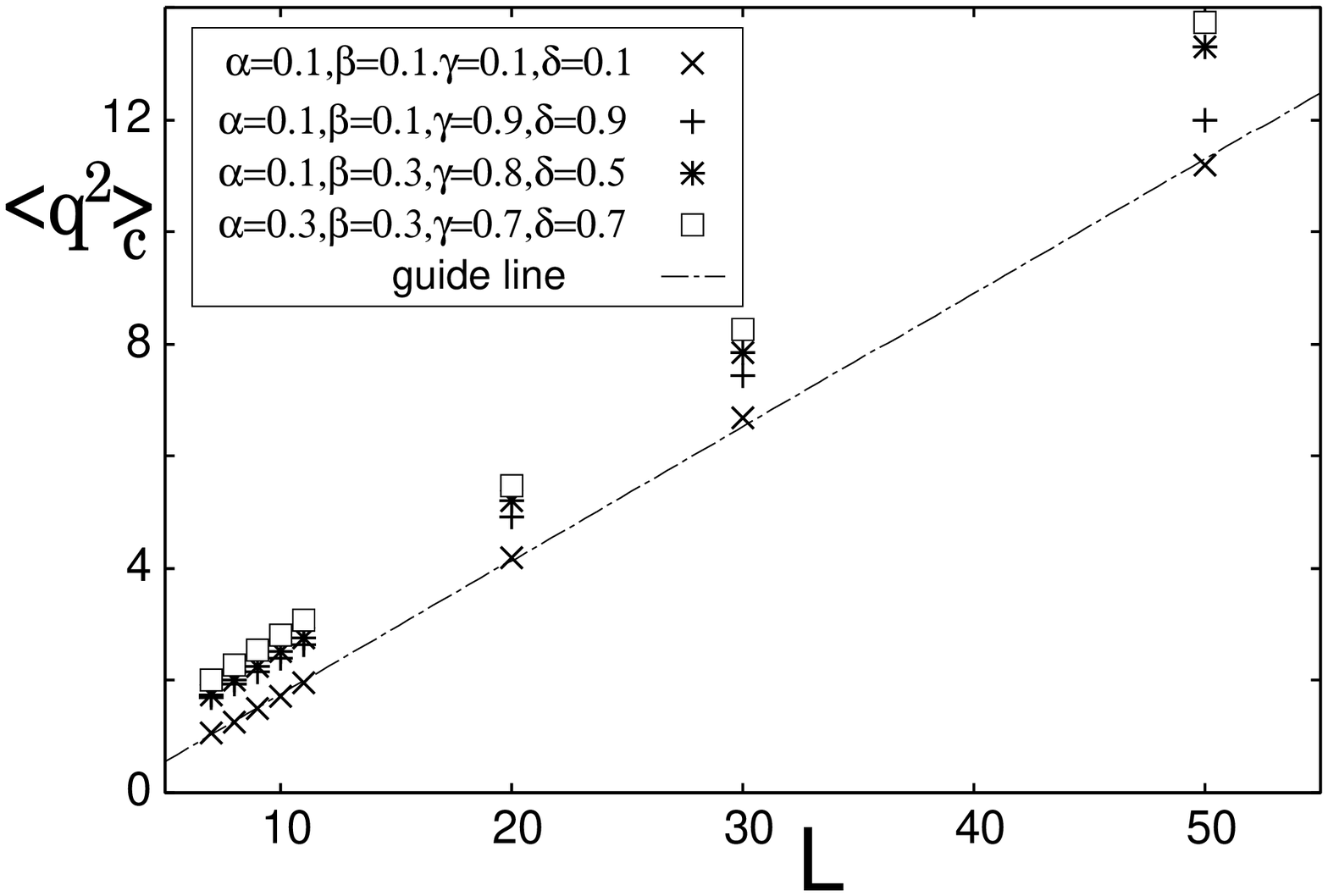}
\includegraphics[scale=0.35]{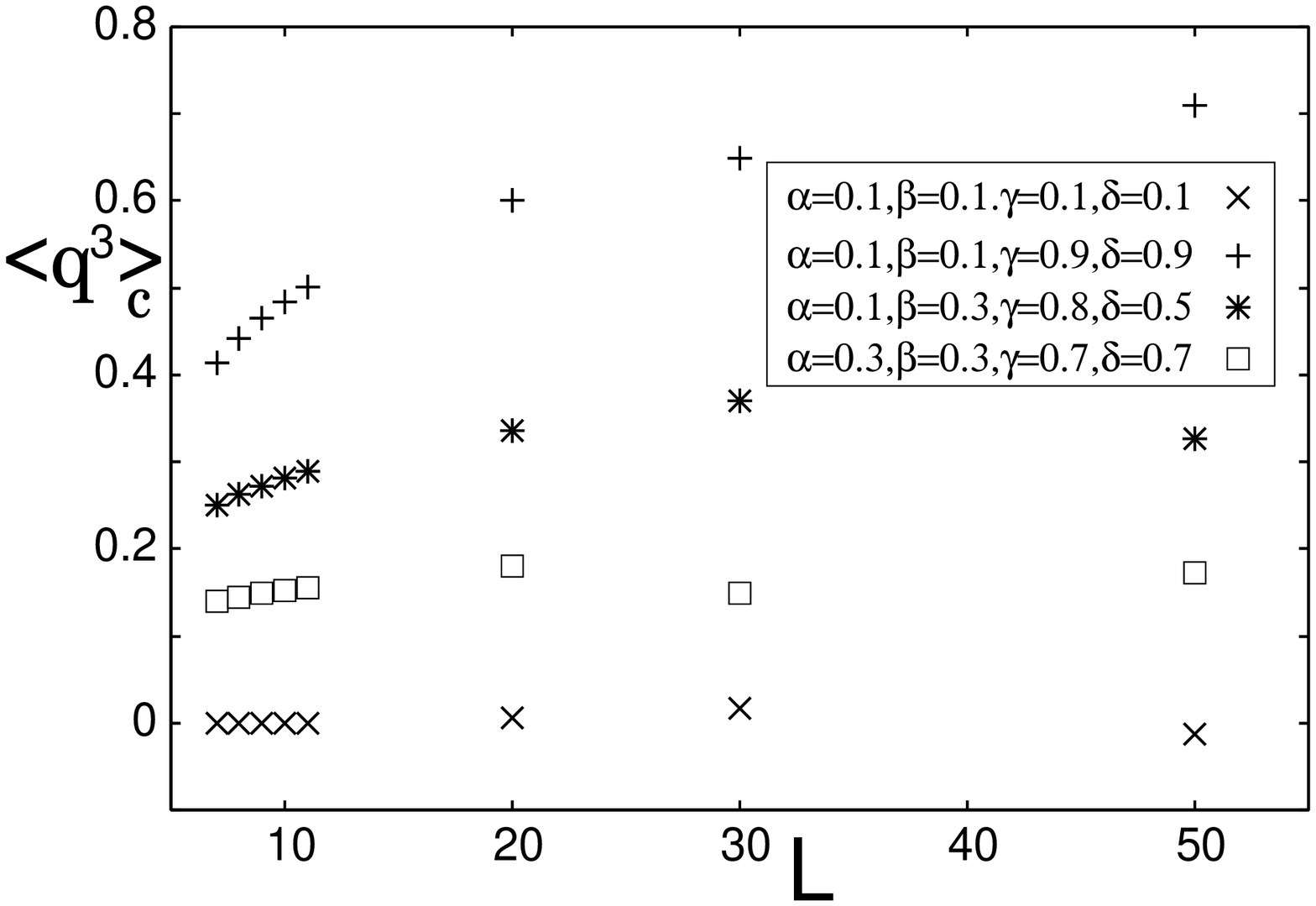}
\includegraphics[scale=0.35]{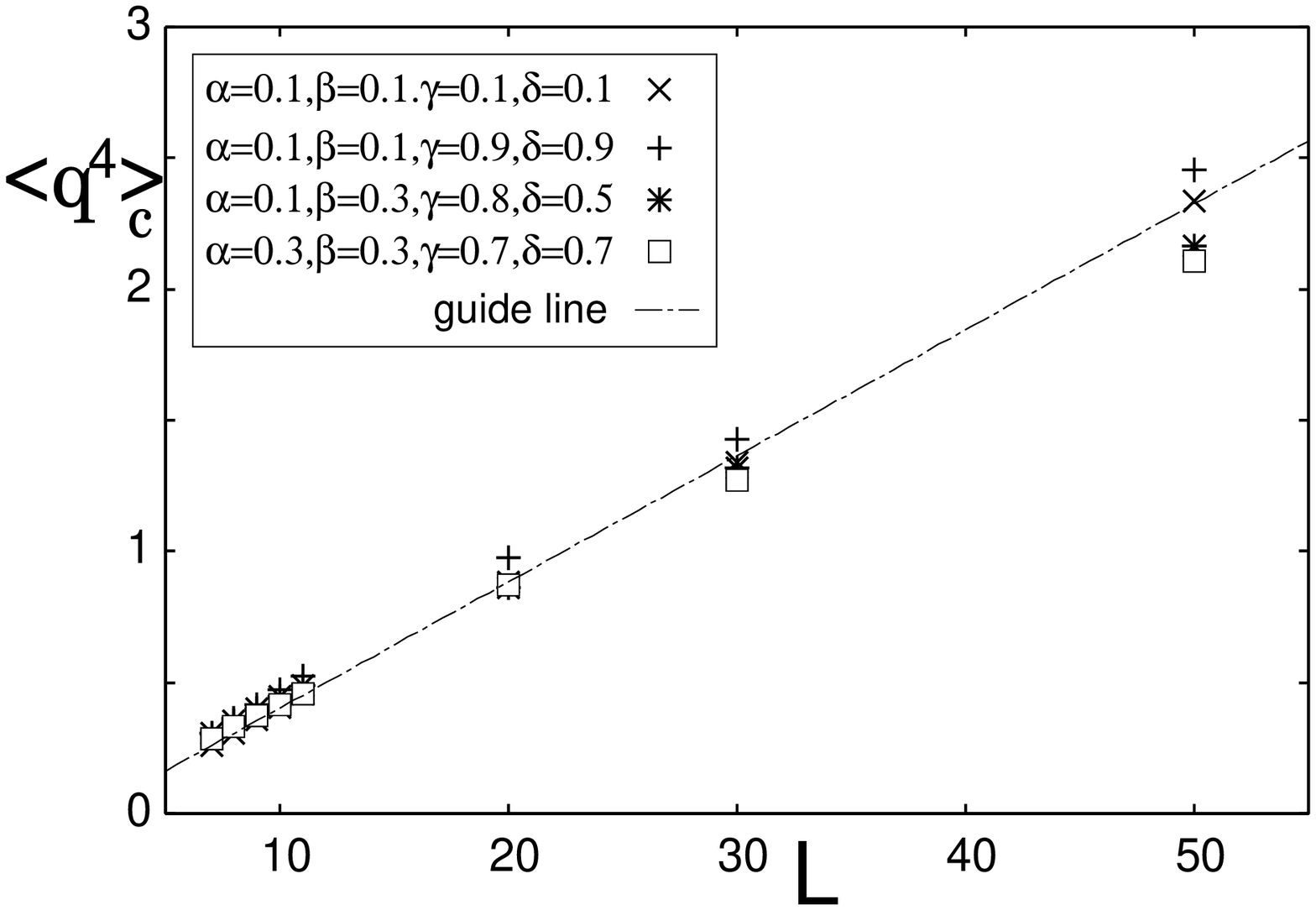}
\includegraphics[scale=0.35]{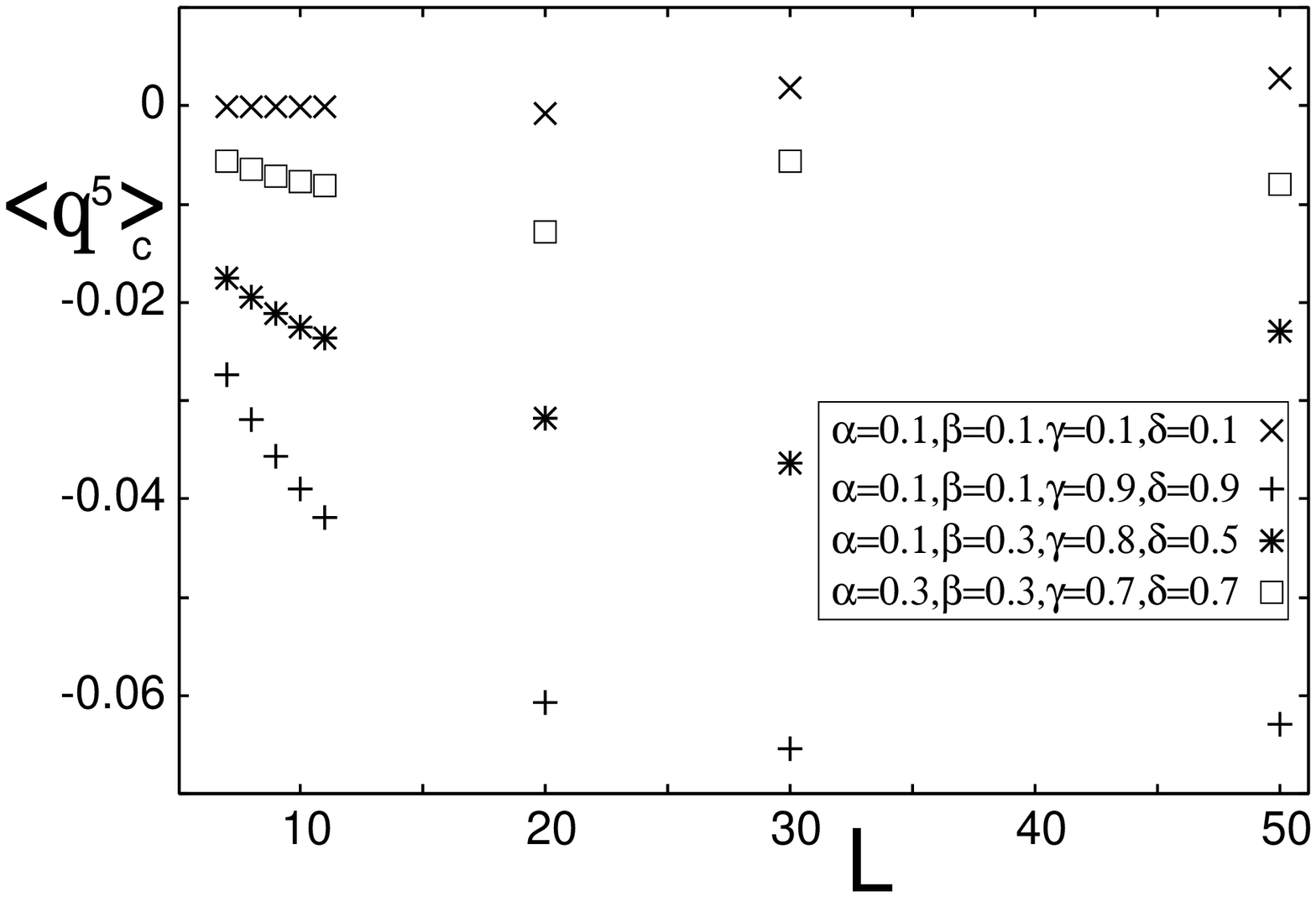}
\includegraphics[scale=0.35]{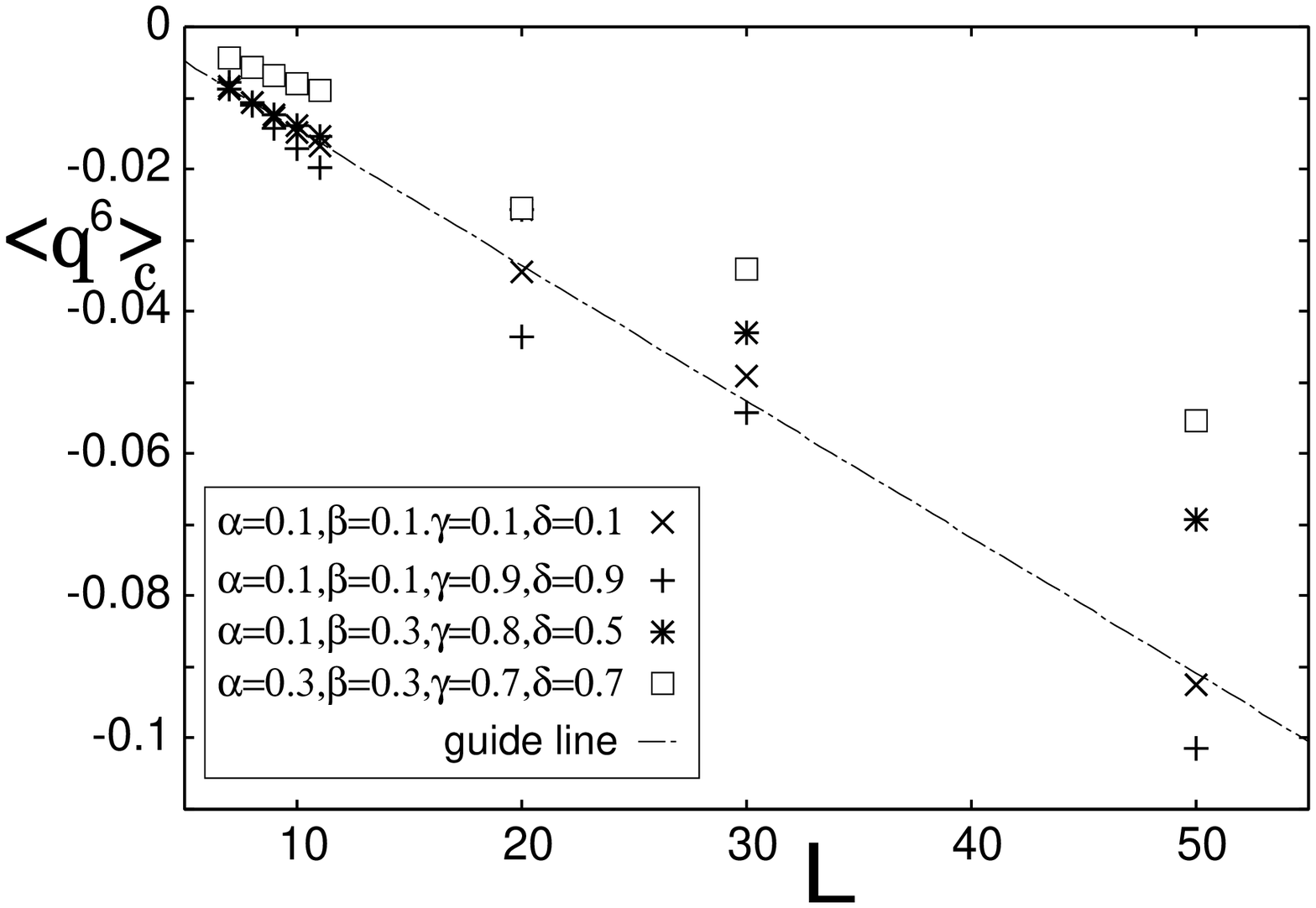}
\caption{2nd- through 6th-order cumulants predicted from the fitting of the
 generating functions given by the simulations. The data of
 $7\leq L\leq11$ are given by the largest eigenvalue and the other data are given
 by the Monte Carlo methods. The data are obtained using the parameters
$(\a,\b,\g,\d)=(0.1,0.1,0.1,0.1)$, $(0.1,0.1,0.9,0.9)$,
 $(0.1,0.3,0.8,0.5)$, and $(0.3,0.3,0.7,0.7)$. }
\label{fig3}
\end{center}
\end{figure}

The system size dependence of some low-order cumulants is shown in Fig. \ref{fig3},
where the data of $7\le L\le 11$ are obtained from the largest
eigenvalues and those of the larger sizes from the Monte Carlo simulations.
The boundary parameters are $(\a,\b,\g,\d)=(0.1,0.1,0.1,0.1)$,
$(0.1,0.1,0.9,0.9)$, $(0.1,0.3,0.8,0.5)$, and $(0.3,0.3,0.7,0.7)$. 
In every case, the cumulants of
even order are proportional to the system size but those of odd order are
not. 
Thus, the large deviation function cannot be estimated by the
mean and the variance only; its characterization needs higher-order
statistical quantities.
However, we must note that the present fitting is not very reliable, because 
the result depends on the range of data and on the function to fit with.
To obtain more accurate results for the higher-order cumulants, we need a wider
range of data.
In particular, the cumulants of odd order have relatively small magnitudes
and fitting errors are significantly large.
Thus, we could not determine how these cumulants really change with the system
size.

From our results, we conjecture that the generating function split
into three parts as
\begin{equation}
\label{threep}
 \m_A(\l)=\bra q_A \ket \l+\m_{A,\mathrm{odd}}(\l)+\m_{A,\mathrm{even}}(\l).
\end{equation}
The first term in the rhs represents the mean current, which does not depend
on the system size for the SSEP and is proportional to the system size
for the ASEP.
The second term is the even part, which is proportional to the system size $L$. 
We are not certain how the odd part behaves.

\begin{figure}
\begin{center}
\includegraphics[scale=0.5]{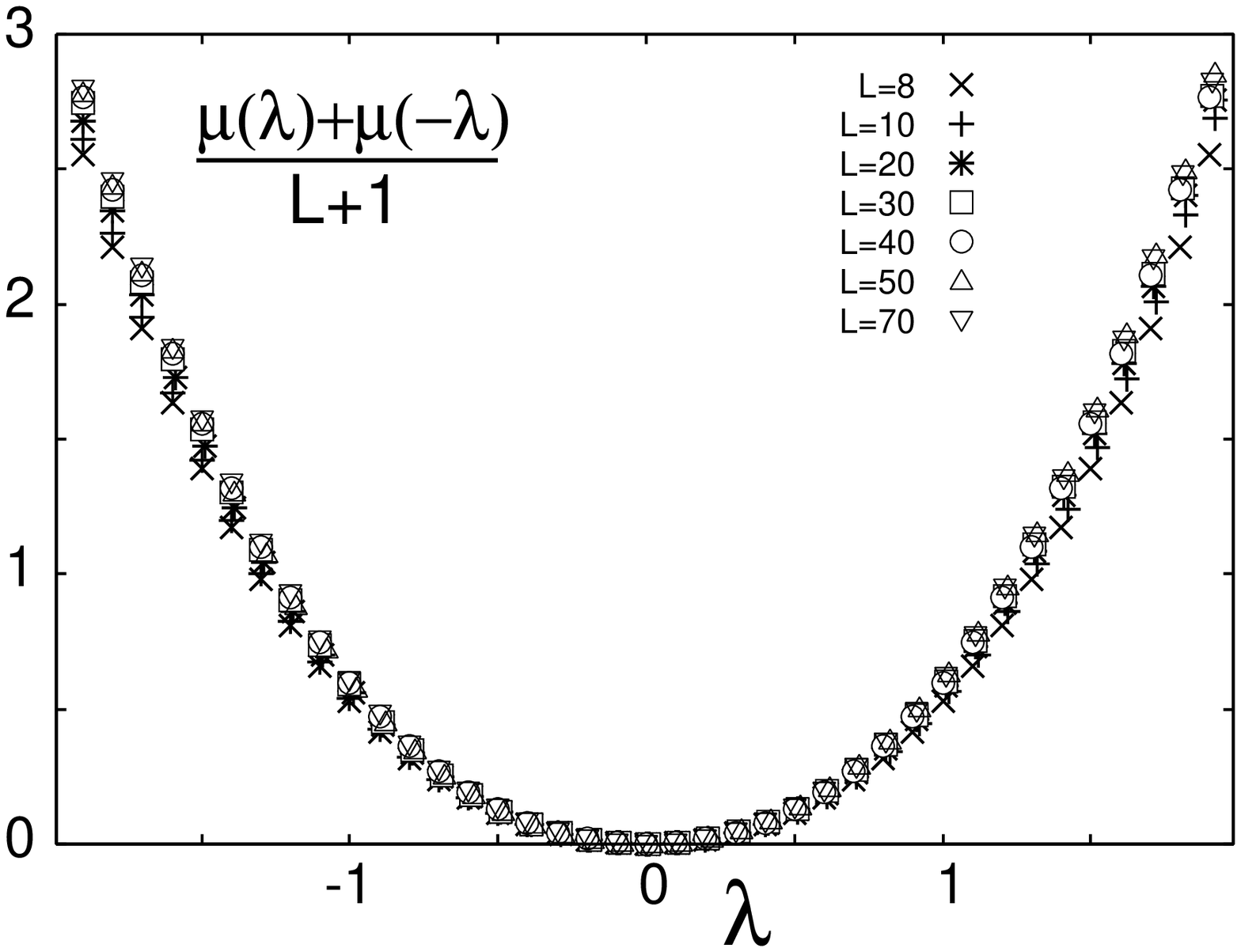}
\caption{The function $(\m_A(\l)+\m_A(-\l))/(L+1)$ is plotted with
 the system sizes of $L=8(\times)$, $10(+)$, $20(\ast)$, $30(\Box)$,
 $40(\bigcirc)$, $50(\triangle)$, and $60(\triangledown)$.}
\label{fig4}
\end{center}
\end{figure}

We show the data of $(\m_A(\l)+\m_A(-\l))/(L+1)=\frac{2}{L+1}\m_{A,\mathrm{even}}(\l)$ when
$L=8$, $10$, $20$, $30$, $40$, $50$, and $70$ in Fig. \ref{fig4}, 
which shows data collapse.
This supports the conjecture that the even part of the generating
function is proportional to the system size.

\subsection{ASEP with the open boundary conditions}

\begin{figure}
\begin{center}
\includegraphics[scale=0.5]{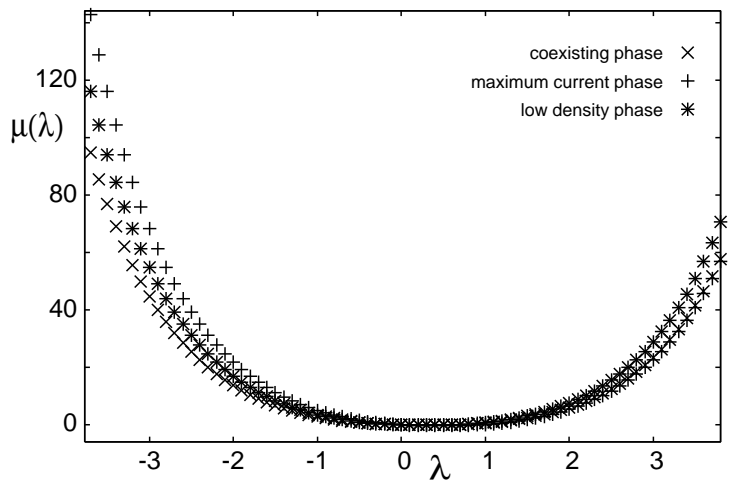}
\caption{Generating function $\m_A(\l)$ in the ASEP when $L=10$ and
 $p_l=0.5$. Crosses, pluses, and stars correspond to the coexisting phase
 ($(\a,\b,\g,\d)=(0.1,0.1,0.1,0.1)$), the maximum-current phase 
($(0.9,0.9,0.1,0.1)$), and
 the low-density phase ($(0.1,0.9,0.9,0.1)$), respectively.}
\label{fig5}
\end{center}
\end{figure}

\begin{figure}
\begin{center}
\includegraphics[scale=0.5]{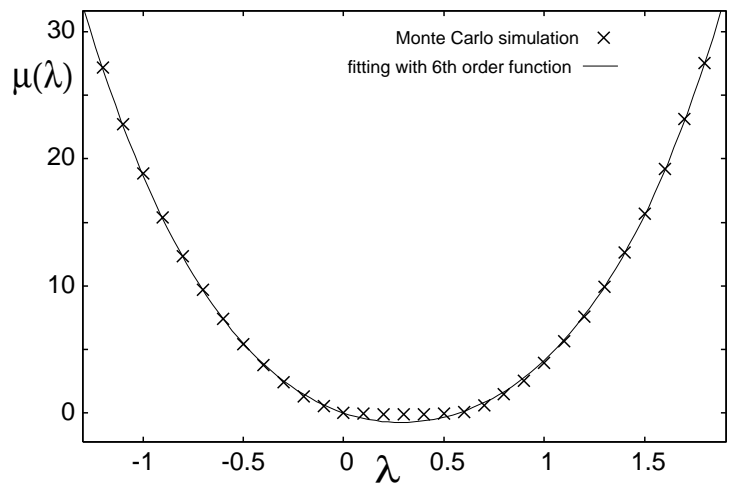}
\caption{Fitting of the generating function $\m_A(\l)$ with 6th-order
 function in the maximum-current phase. We may fit $\m_A(\l)$ well in the region
 outside the minimum, and this can be seen in other phases.}
\label{fig6}
\end{center}
\end{figure}

Now, we present the results on the open boundary ASEP, where the
transition rates satisfy $p_l<p_r=1$, and observe whether there are any differences
among phases.  
Because the high-density phase is equivalent to the low-density phase due to
the particle-hole symmetry, we examine the remaining two phases (the low-density phase
and the maximum-current phase) and the
coexisting state of the low- and high-density phases.
Figure \ref{fig5} illustrates the generating function $\m_A(\l)$ in each
phase of the ASEP with system size $L=10$ and $p_l=0.5$.
We note that no corresponding analytical or numerical results are
indicated 
in the literature.
It is remarkable that any qualitative differences are not observed
among the three states.
As in the SSEP, 
we have attempted to fit the data with the 6th-order polynomial, and 
the result for $L=50$, $p_l=0.5$, and $(\a,\b,\g,\d)=(0.9,0.9,0.1,0.1)$ 
is shown in Fig. \ref{fig6}.
The fitting seems to work well if a certain neighborhood of the minimum is excluded.
We confirmed this property also in other phases.  This flattening of
the minimum has been reported also in a system with  
periodic boundary condition\cite{Pro09}.
Clearly, the generating function cannot be assumed to be quadratic.

\begin{figure}
\begin{center}
\includegraphics[scale=0.5]{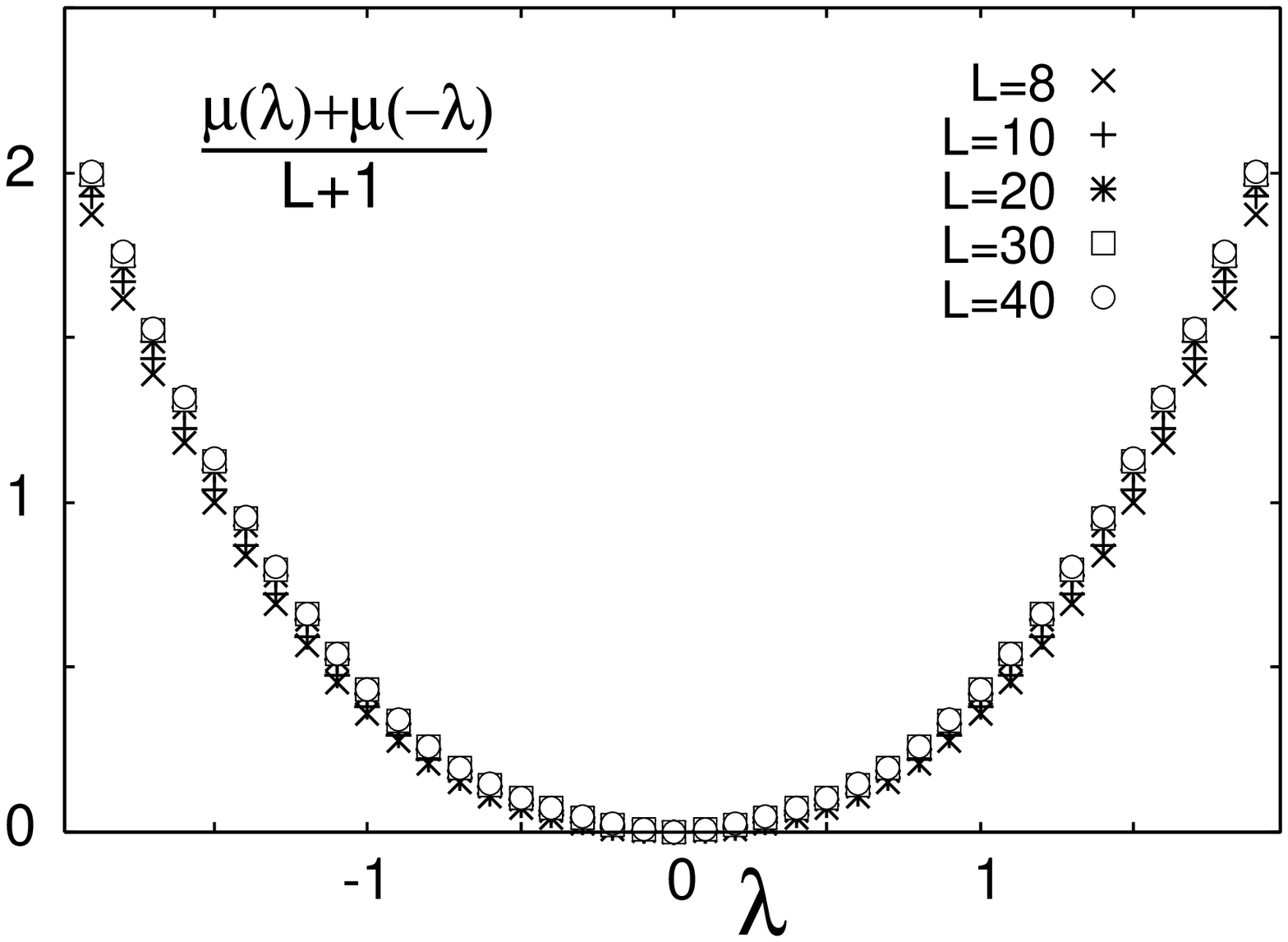}
\includegraphics[scale=0.5]{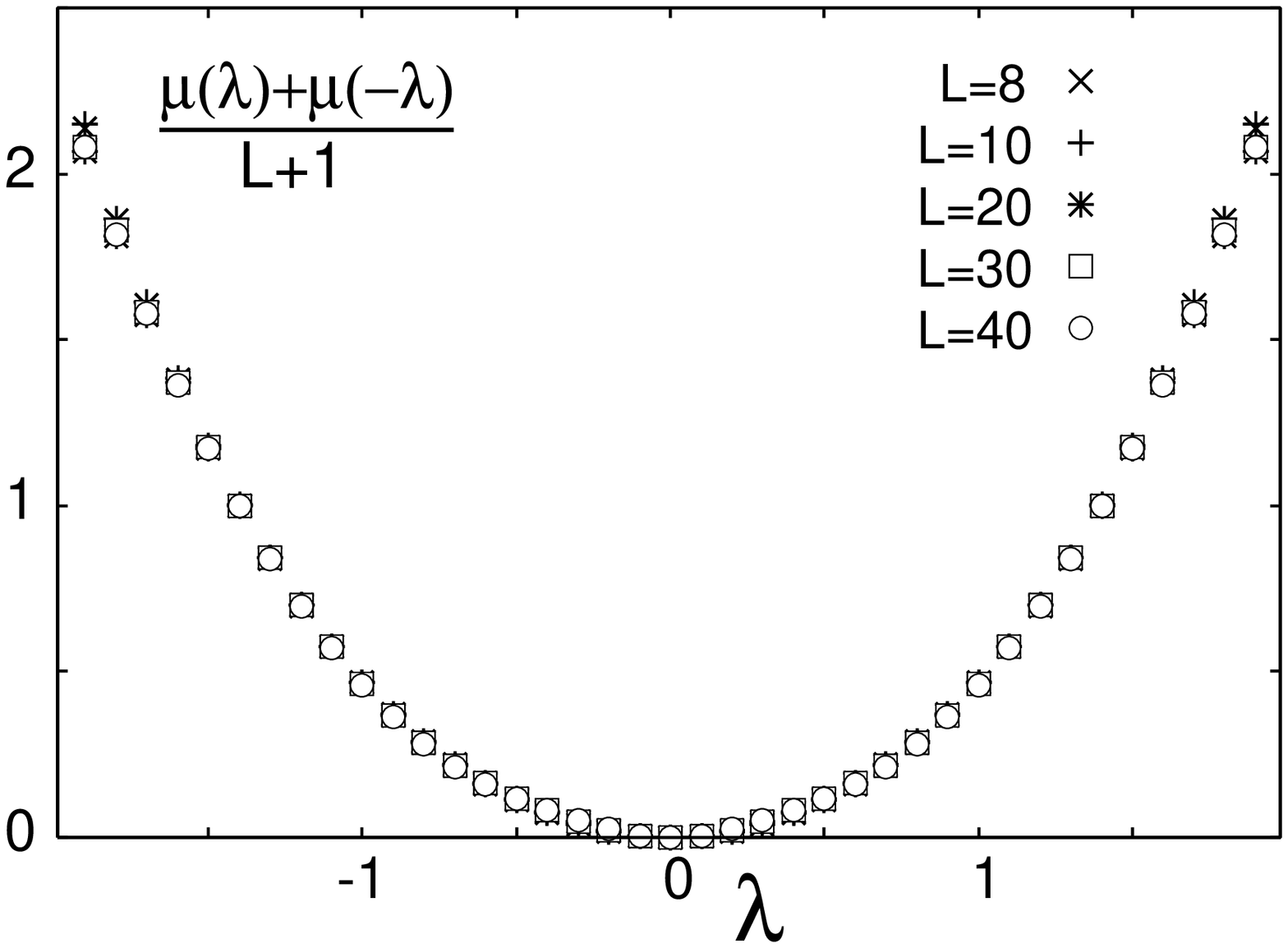}
\includegraphics[scale=0.5]{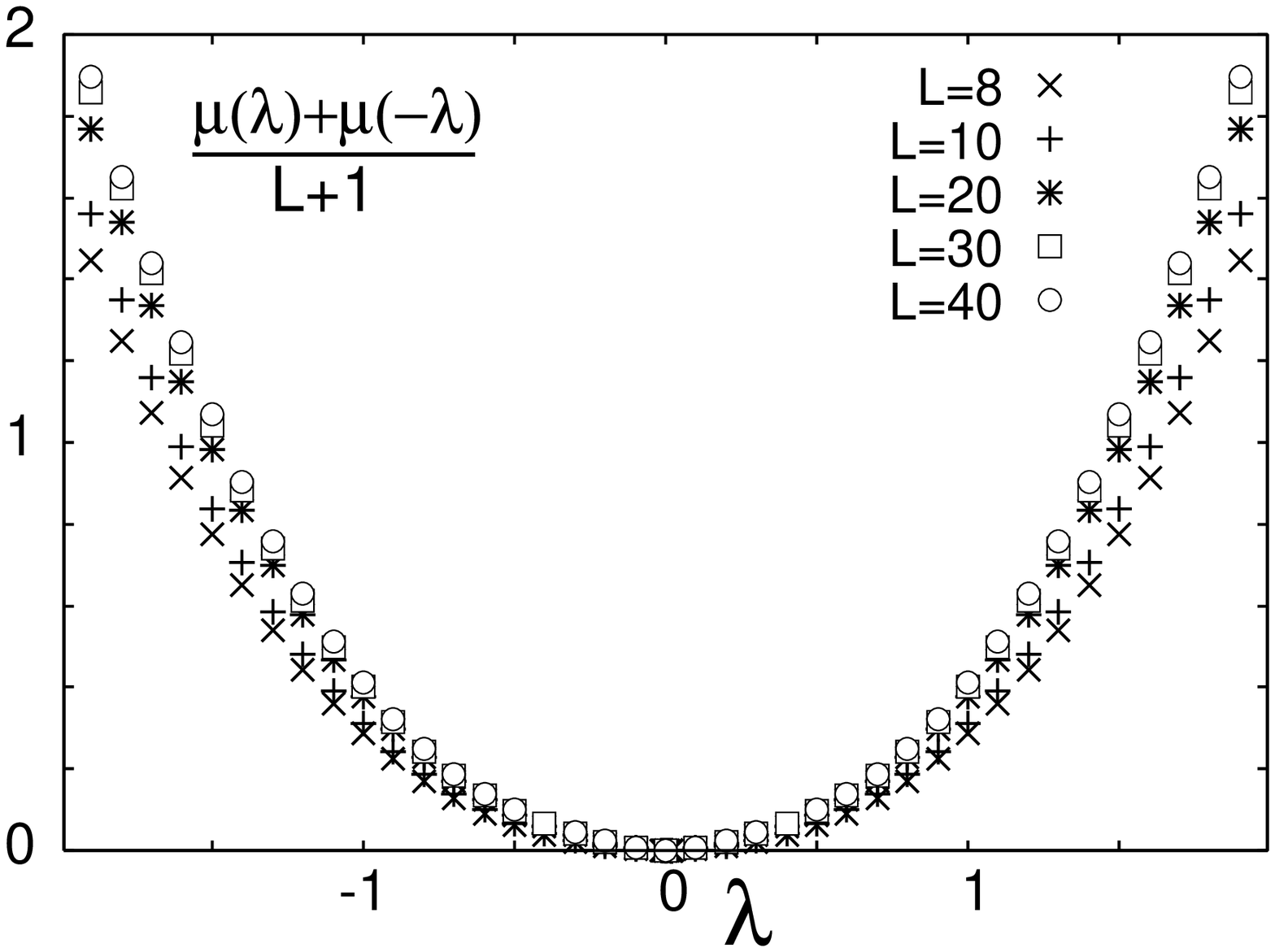}
\caption{The function $(\m_A(\l)+\m_A(-\l))/(L+1)$ is plotted with the
 system sizes of $L=8(\times)$, $10(+)$, $20(\ast)$, $30(\Box)$, and $40(\bigcirc)$,
 in the low-density phase (top), maximum-current phase (middle) and
 coexisting phase (bottom).}
\label{fig7}
\end{center}
\end{figure}

Similarly to the case of the SSEP, we may divide the generating function
for the ASEP into three parts as in eq. (\ref{threep}).
We illustrate $(\m_A(\l)+\m_A(-\l))/(L+1)$ in Fig. \ref{fig7} for
various system sizes $L=8$, $10$, $20$, $30$, and $40$.
In relatively small systems, 
the maximum current phase shows very good data collapse, while
the degree of data collapse is not so good in other phases.
In every case, however, the data seem to converge in the large-system limit.
Thus, as in the SSEP, the even part of the generating function is
proportional to the system size when it is sufficiently large.

\begin{figure}
\begin{center}
\includegraphics[scale=0.4]{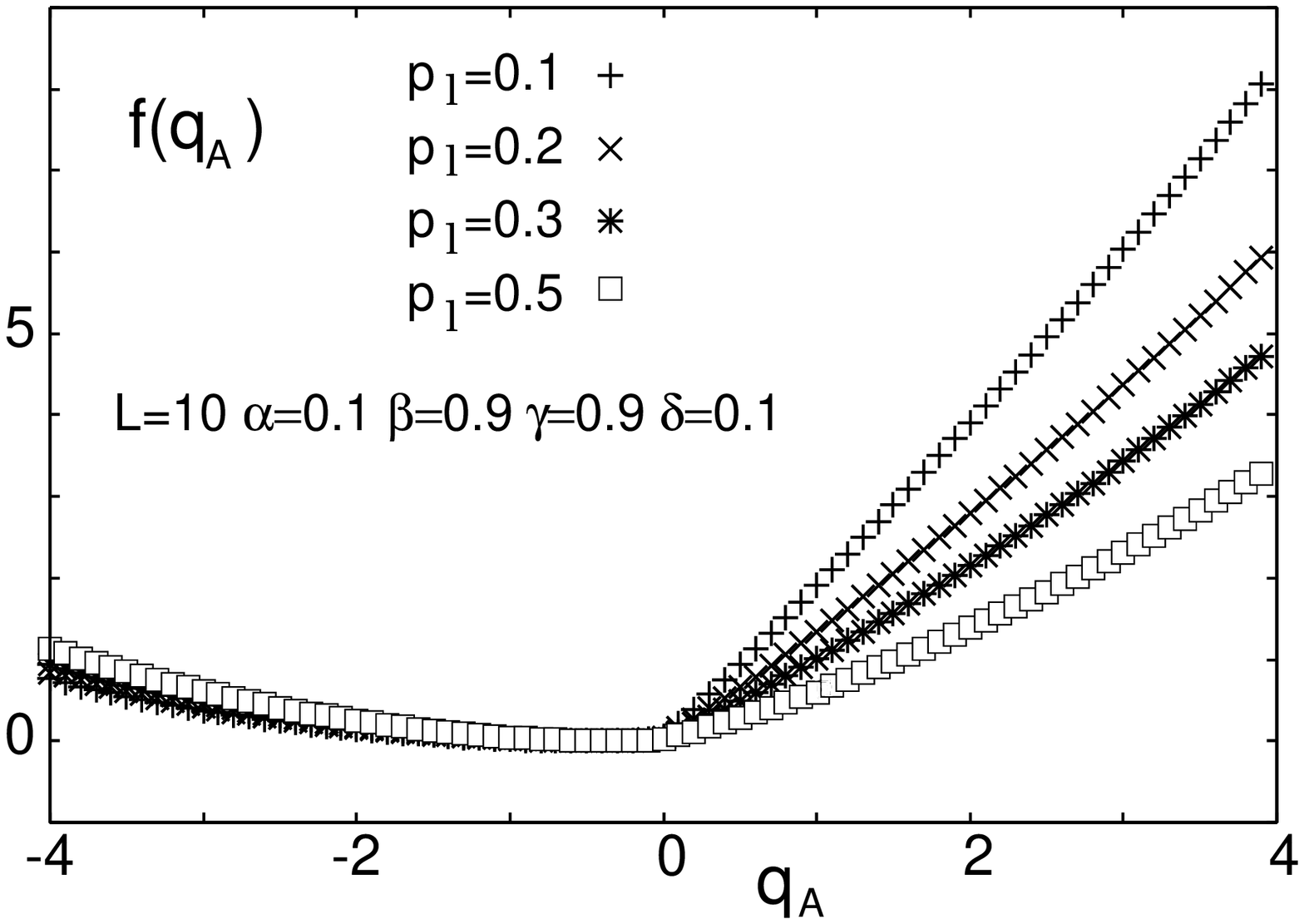}
\includegraphics[scale=0.4]{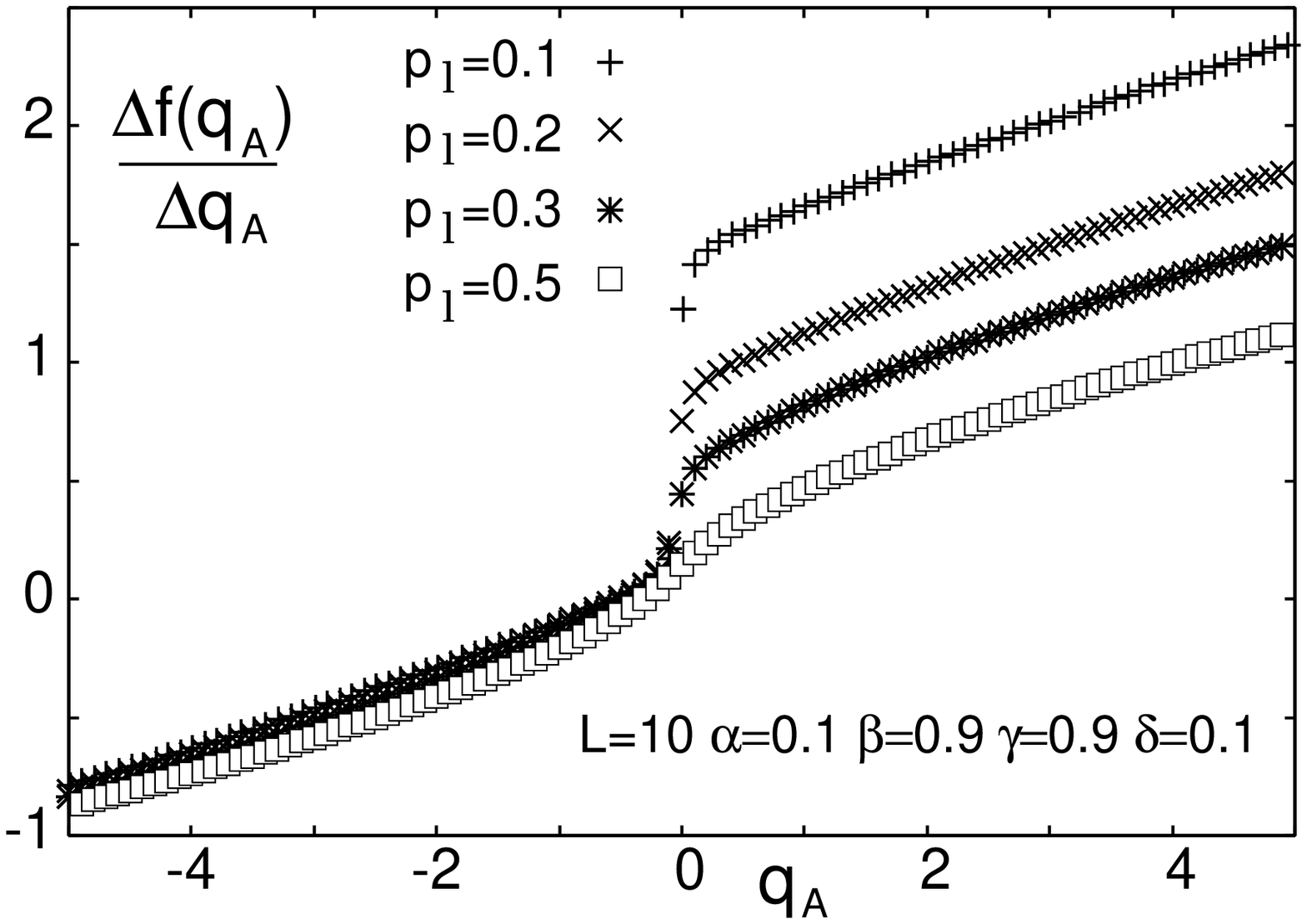}
\caption{Large deviation function given by the Legendre
 transform in the range $-5\leq q_A\leq5$. The parameters are $L=10$,
 $(\a,\b,\g,\d)=(0.1,0.9,0.9,0.1)$ (low-density phase), and
 $p_l=0.1$, $0.2$, $0.3$, and $0.5$. The derivative of the left figure is shown in
 the right figure.}
\label{fig8}
\end{center}
\end{figure}

The large deviation function is obtained from the generating function via 
the Legendre transformation.
Figure \ref{fig8} shows the plot for $L=10$,
$(\a,\b,\g,\d)=(0.1,0.9,0.9,0.1)$, and $p_l=0.1,0.2,0.3,0.5$, 
all of which correspond to the low density phase, in the range $-4\leq q\leq4$.
A blowup near the minimum of $f(q_A)$ is shown on left of Fig. \ref{fig8}.
We note that a cusp appears near $q=0$ and its sharpness increases as
$p_l$ decreases.  This is confirmed by the plot of the derivative of
the large deviation function shown on right of Fig. \ref{fig8}.
The cusp means that the probability of positive current rapidly
decreases with $L$.
Such a cusp is seen also in other phases.
In the limit $p_l\rightarrow 0$, the particles do not move leftward except at
the boundaries. 
Thus, the large deviation function is expected to diverge for $q_A>0$ in
the limit.
This is why the cusp is generated.

\subsection{Comparing the two currents}

Thus far, we have presented the simulation results for the total currents.
In this subsection, we compare the characteristic functions of
$q_A$ and $q_B$.
First, we introduce the
dimensionless current $q^N_X$ defined by
\begin{equation}
 q^N_X=\frac{q_X}{\bra q_X\ket},
\end{equation}
where $\bra q_X\ket$ is the expectation,
and then we write the generating function $\m^N_X(\l)$ of the current $q_X^N$.
We compare $\m^N_A(\l)$ and $\m^N_B(\l)$ in the simulations.

The dimensionless current defined here is effective when $\bra q_X\ket$
is known.
Fortunately, the steady current is exactly calculated for both the
SSEP\cite{Der01} and the ASEP\cite{Uch00} as
\begin{equation}
\label{qbssep}
 \bra q_B\ket_{SSEP}=-\frac{\frac{\d}{\b+\d}
  -\frac{\a}{\g+\a}}{L+\frac{1}{\a+\g}+\frac{1}{\b+\d}-1},
\end{equation}
for the SSEP, and
\begin{equation}
 \bra q_B\ket_{ASEP}=\left(
\begin{array}{cc}
(p_l-1)\r_a(1-\r_a), & \mbox{low-density and coexisting phases} \\
(p_l-1)\r_b(1-\r_b), & \mbox{high-density phase} \\
\frac{p_l-1}{4} & \mbox{maximum-current phase} 
\end{array}\right. 
\end{equation}
for the ASEP with $p_l<p_r=1$, where $\r_a$ and $\r_b$ are given as
\begin{equation}
 \r_a=\frac{1}{1+a} \qquad a=\frac{1-p_l-\a+\g+\sqrt{(1-p_l-\a+\g)^2+4\a\g}}{2\a} 
\end{equation}
and
\begin{equation}
 \r_b=\frac{b}{1+b} \qquad b=\frac{1-p_l-\d+\b+\sqrt{(1-p_l-\d+\b)^2+4\b\d}}{2\b}.
\end{equation}
Henceforth, we omit the suffix `SSEP' or `ASEP' from the steady current
$\bra q_X\ket$, for it can be easily understood from the context.

\begin{figure}
\begin{center}
\includegraphics[scale=0.4]{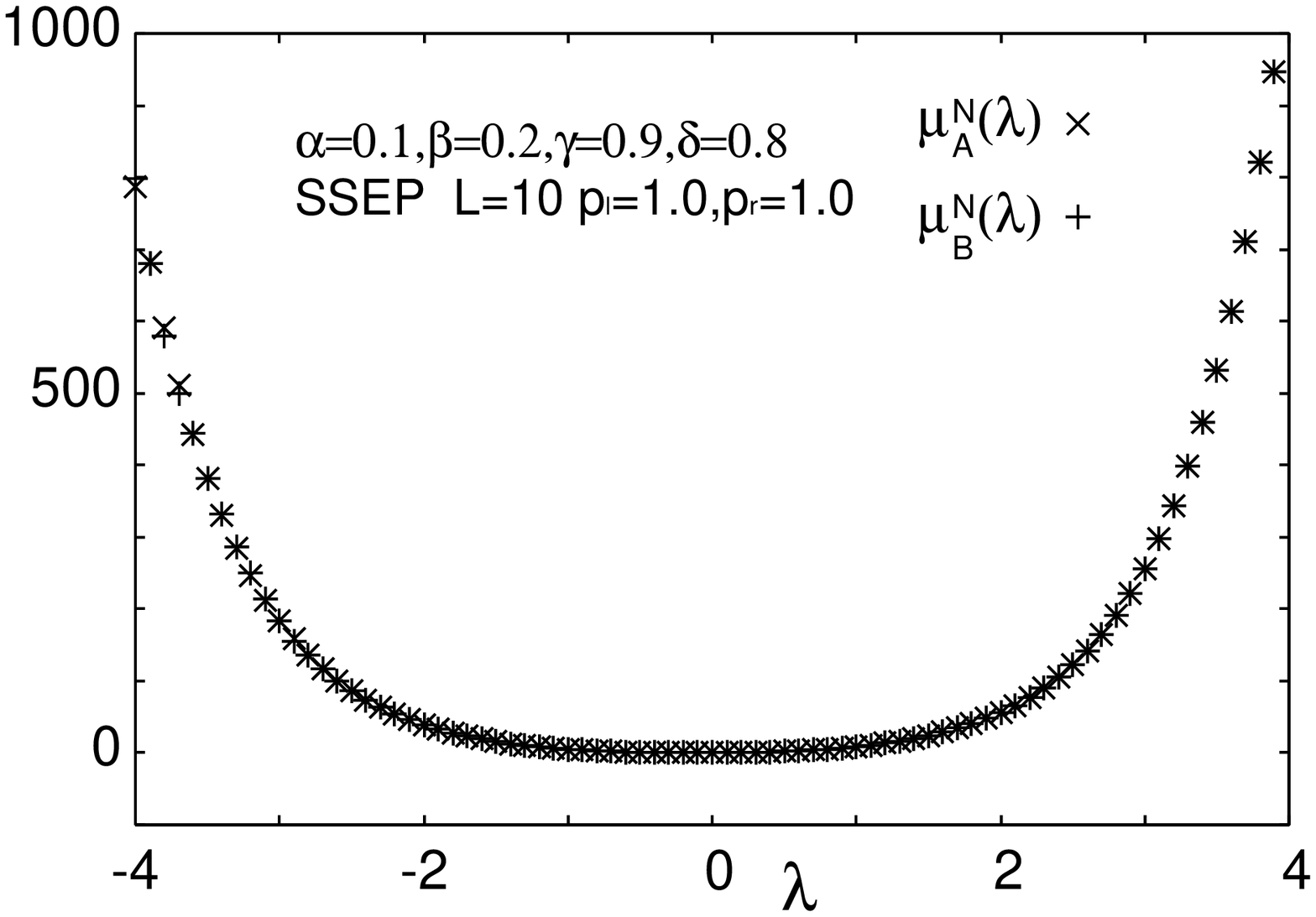}
\includegraphics[scale=0.4]{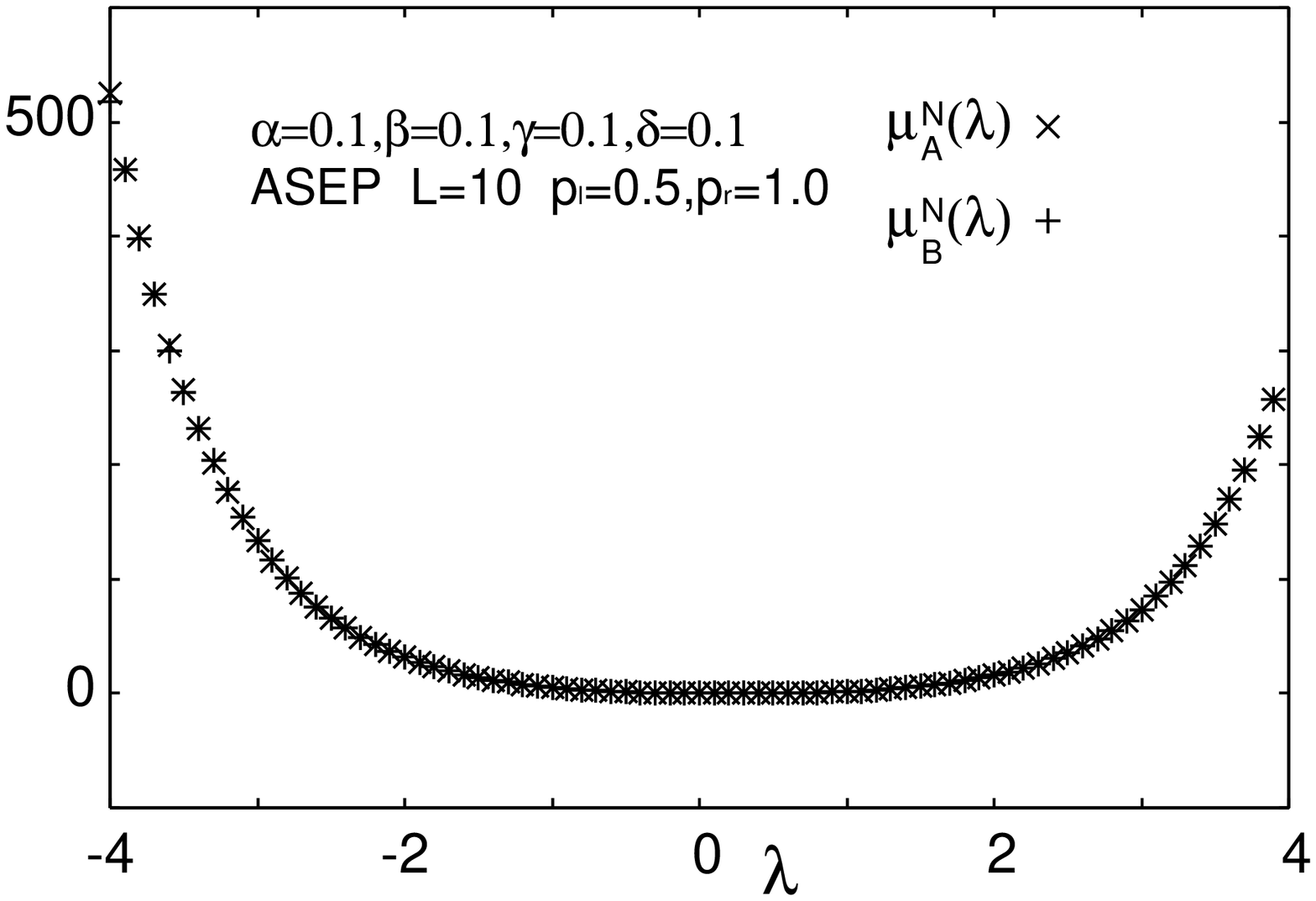}
\includegraphics[scale=0.4]{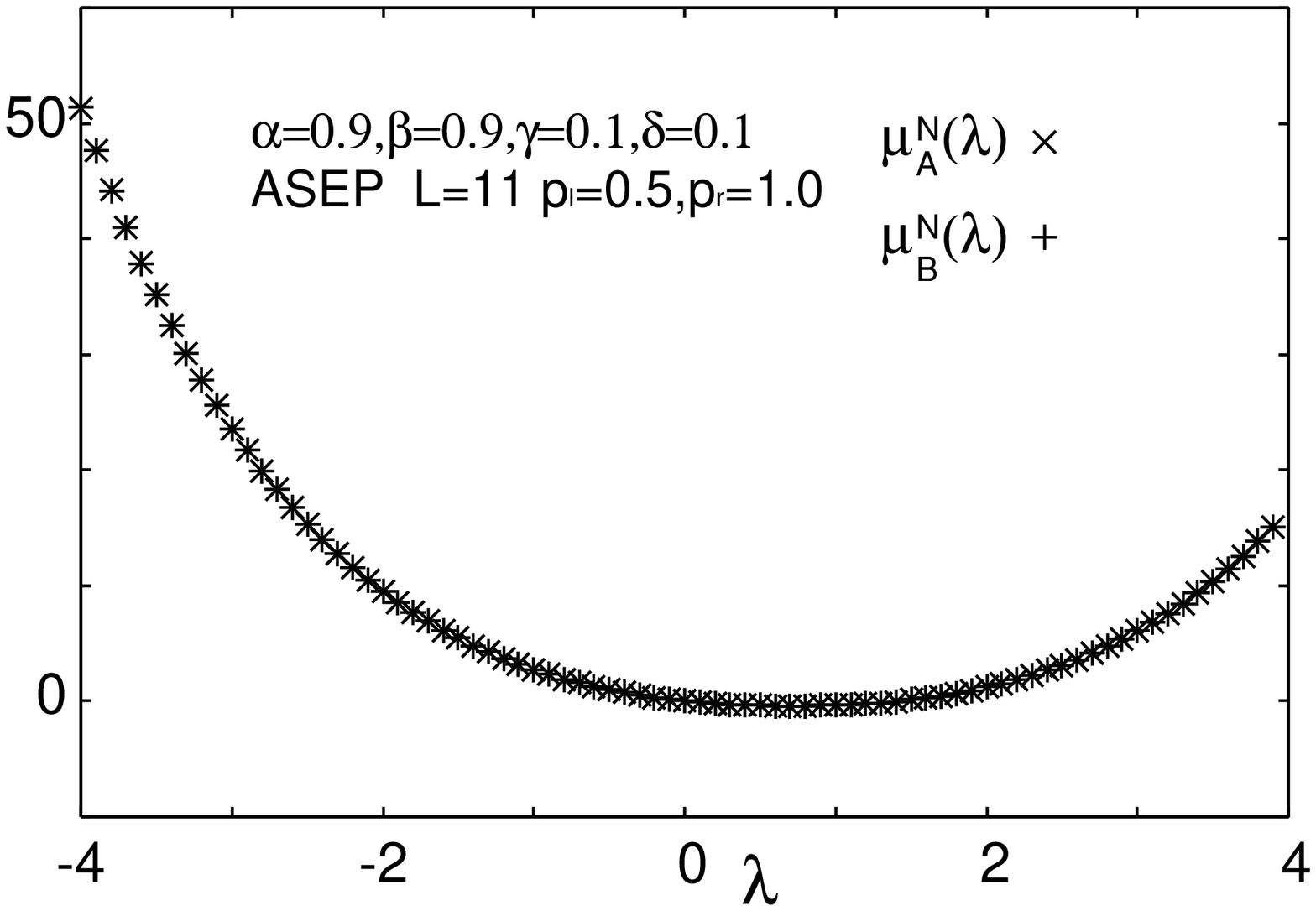}
\caption{Plot of $\m^N_A(\l)$($\times$) and $\m^N_B(\l)$($+$) for
 three sets of parameters. We observe a clear coincidence of
 $\m^N_A(\l)$ and $\m^N_B(\l)$.}
\label{fig9}
\end{center}
\end{figure}

The simulation results from the method of largest eigenvalues are 
shown in Fig. \ref{fig9}.
It is clear that $\m^N_A(\l)$ and $\m^N_B(\l)$ coincide with each other.

\begin{figure}
\begin{center}
\includegraphics[scale=0.5]{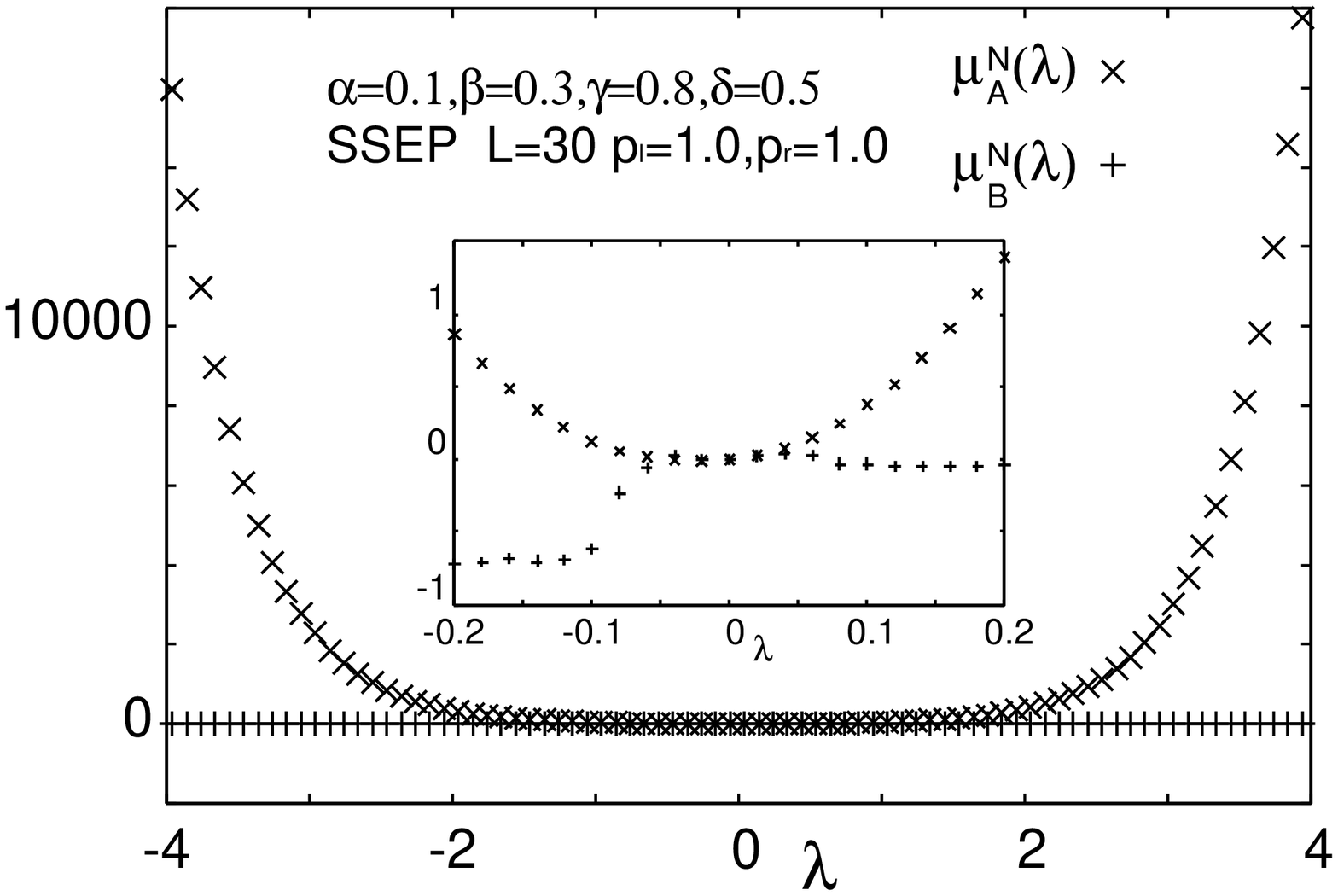}
\includegraphics[scale=0.5]{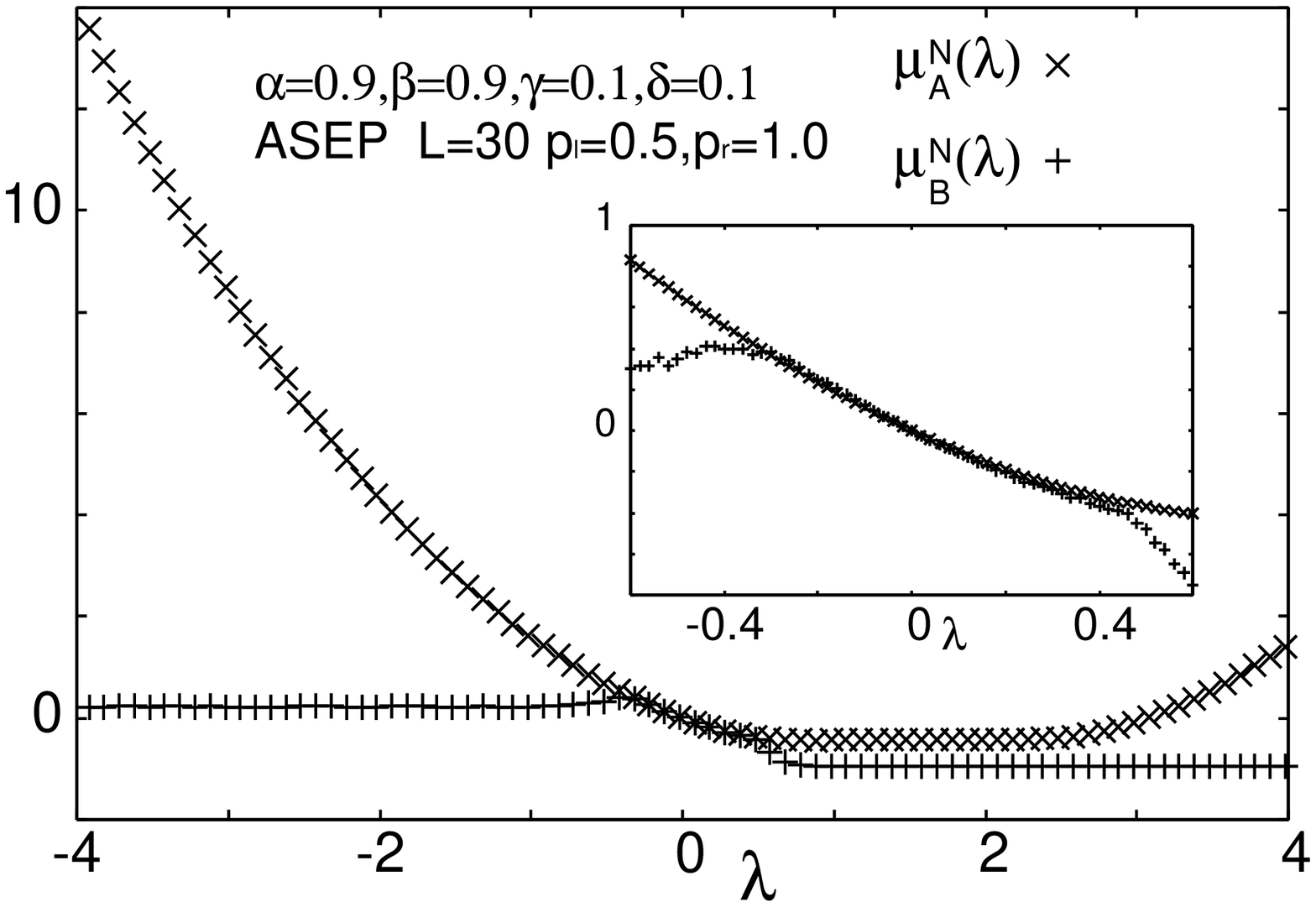}
\caption{Plot of $\m^N_A(\l)$ and $\m^N_B(\l)$ for the SSEP and the
 ASEP. The inset shows the extended plot of the coinciding area.}
\label{fig10}
\end{center}
\end{figure}

\begin{figure}
\begin{center}
\includegraphics[scale=0.5]{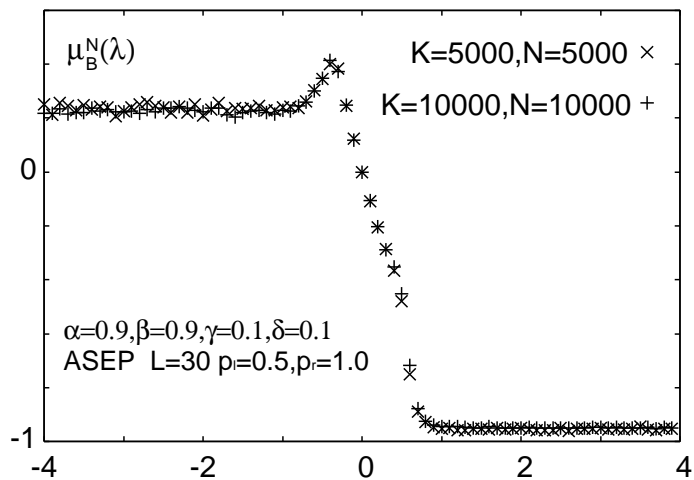}
\caption{Plot of $\m^N_B(\l)$ for different time
 step $K$ and clone number $N$ values.}
\label{fig11}
\end{center}
\end{figure}

However, the two large deviation functions show differences in the Monte Carlo
simulation, as shown in Fig. \ref{fig10}.  They agree with each other
only when $|\l|$ is very small and differ greatly outside the region.
This difference means that $\m^N_B(\l)$ is not appropriately calculated 
in the Monte Carlo simulation.  One reason for this is the lack of statistics.
Because $q_B$ counts only the flux at the left boundary, the number of such
events is much smaller than that in the case of $q_A$.  Another possibility is 
that the number of clones is not sufficiently large.  However, no significant change
has been seen in a simulation where we doubled the number of clones and
the time duration, as seen in Fig. \ref{fig11}.

Thus, the difference between the two currents can affect results of the Monte 
Carlo simulations, though they represent essentially the same physical 
quantity.
The use of the total current $q_A$ has an advantage in this respect.

\subsection{Symmetry relations}

\begin{figure}
\begin{center}
\includegraphics[scale=0.5]{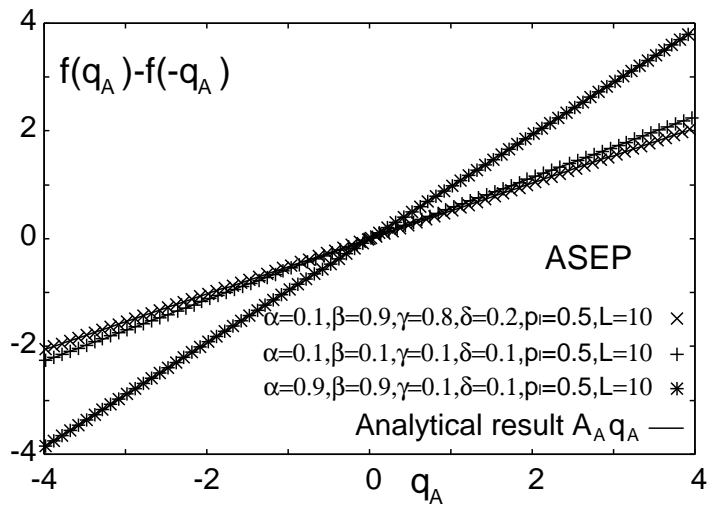}
\caption{Plot of $f(q_A)-f(-q_A)$ from the direct evaluation results
 represented by $\times,+,$ and $\ast$, and that of 
 $A_Aq_A$ from the analytical result obtained using eq. (\ref{aaasep}), 
 which is shown by the solid line.}
\label{fig12}
\end{center}
\end{figure}

The symmetry relation
\begin{equation}
\label{symrel}
 f(q_A)-f(-q_A)=Aq_A
\end{equation}
should hold in the SEP.
First, we show the results for small system sizes obtained
from the direct evaluation of the largest eigenvalue.
We plot $f(q_A)-f(-q_A)$
and $A_Aq_A$ with $A_A$ given by eqs. (\ref{abasep}) and (\ref{aaasep}) in Fig. \ref{fig12}.
The parameter sets used are $(\a,\b,\g,\d)=(0.1,0.9,0.8,0.2)$ for
the low-density phase that we denote by $\times$, $(0.1,0.1,0.1,0.1)$ for
the coexisting phase that we denote by $+$, and $(0.9,0.9,0.1,0.1)$ for
the maximum-current phase that we denote by $\ast$, and the
other parameters $L=10$, $p_l=0.5$, and $p_r=1.0$ used are the same.
The solid line shows $A_Aq_A$. 
We see that the symmetric relation is well satisfied in this case.

\begin{figure}
\begin{center}
\includegraphics[scale=0.5]{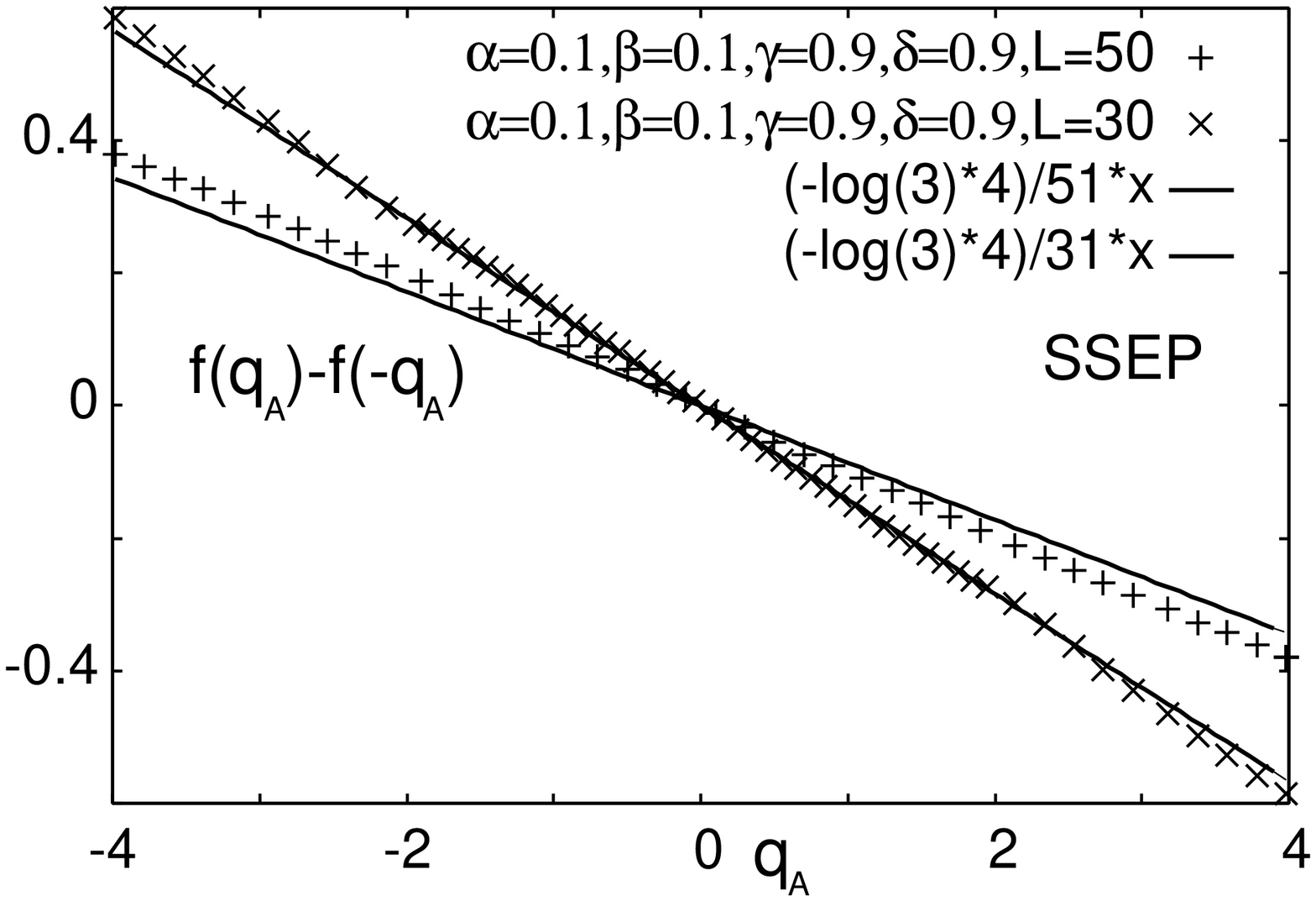}
\includegraphics[scale=0.5]{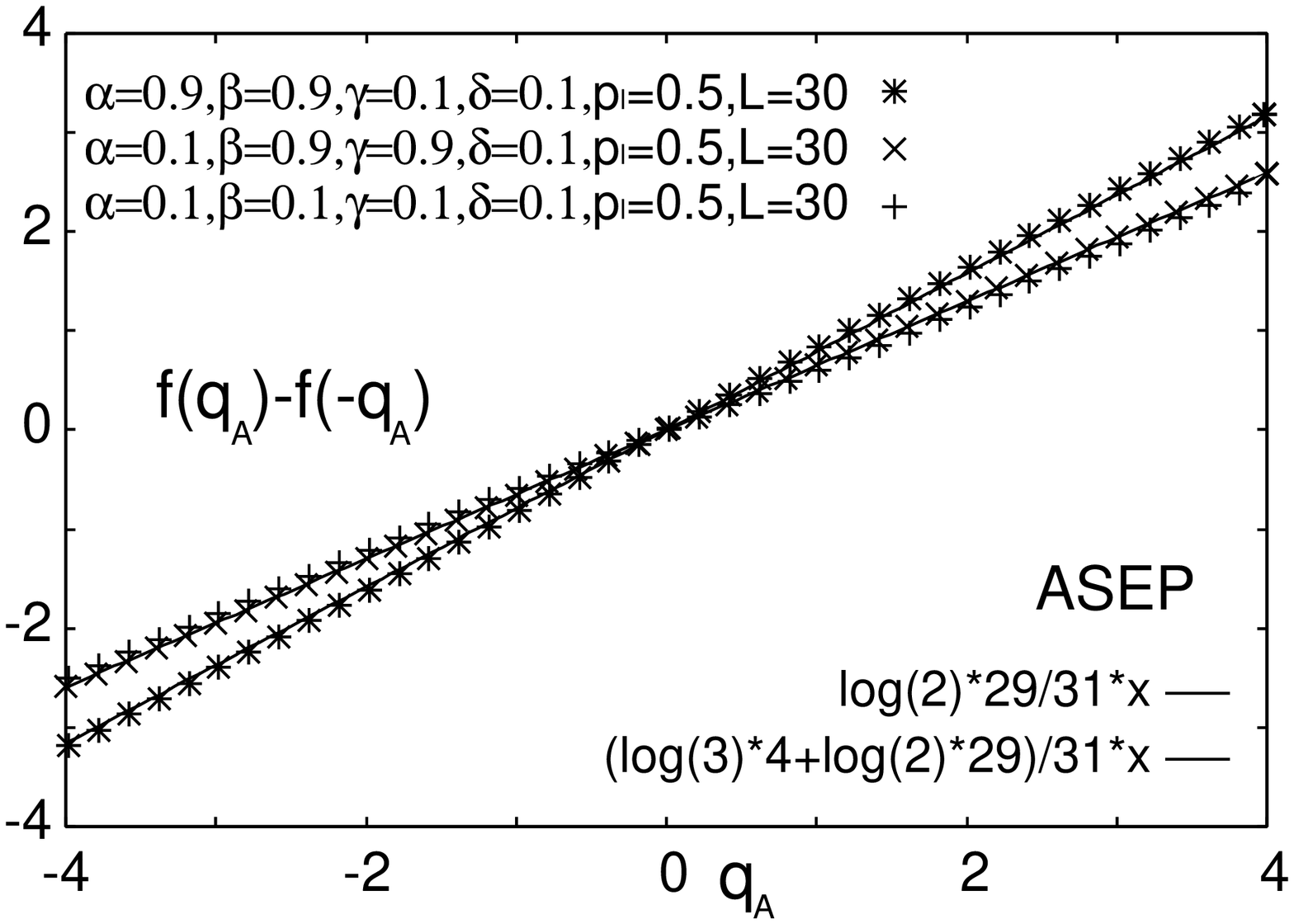}
\includegraphics[scale=0.5]{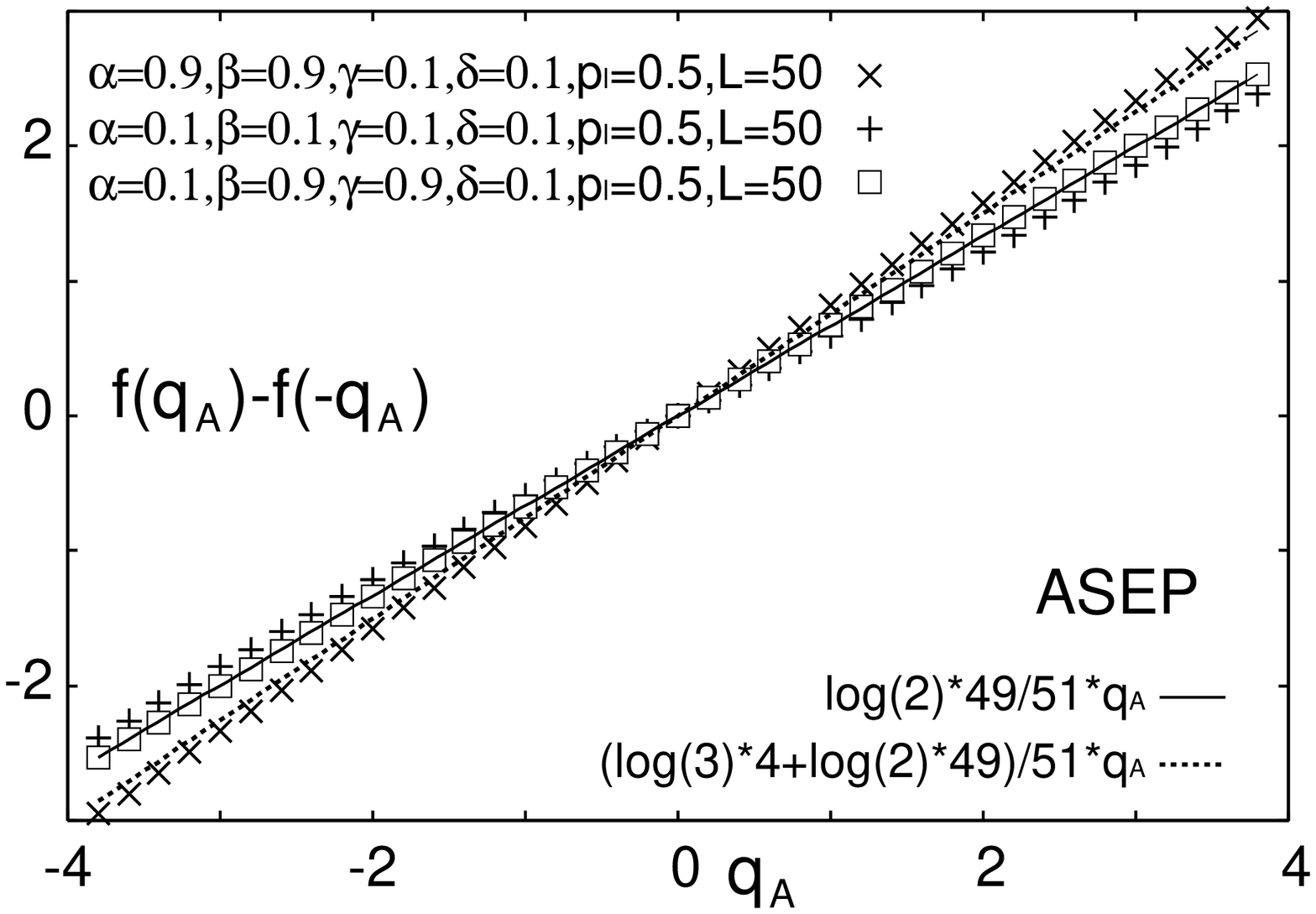}
\caption{Comparison of $f(q_A)-f(-q_A)$ values for the SSEP (top) and
 ASEP (the others). For the SSEP, the system sizes of $L=30$ and $50$ are
 plotted.
For the ASEP, the plot of $L=30$ is given in the second figure and the
 plot of $L=50$ is given in the third figure.
The parameters are given in the inset. The solid and
 dotted lines give the analytical results.}
\label{fig13}
\end{center}
\end{figure}

Next, we consider the results of the Monte Carlo method.  In this case, we used 
the thermodynamic integral to compute the generating function and 
obtained large deviation functions via the Legendre transform.
In Fig. \ref{fig13} we plot $f(q_A)-f(-q_A)$ for the SSEP and the ASEP.
The parameter sets are given in the inset of the figure.
A negligible deviation is observed for the SSEP, while a small deviation
is apparent in the ASEP.

The deviation increases with the system size.
The error may stem from the insufficient number of clones or the
resolution of the data, which affects the precision of the Legendre
transform.
For large system sizes, the number of necessary clones becomes so large that
we could not achieve a higher accuracy.
Thus, deviations in Fig. \ref{fig13} should be interpreted as 
representing the precision and limitation of the present 
Monte Carlo simulations.

\subsection{Convergence problem in the Monte Carlo method}

\begin{figure}
\begin{center}
\includegraphics[scale=0.5]{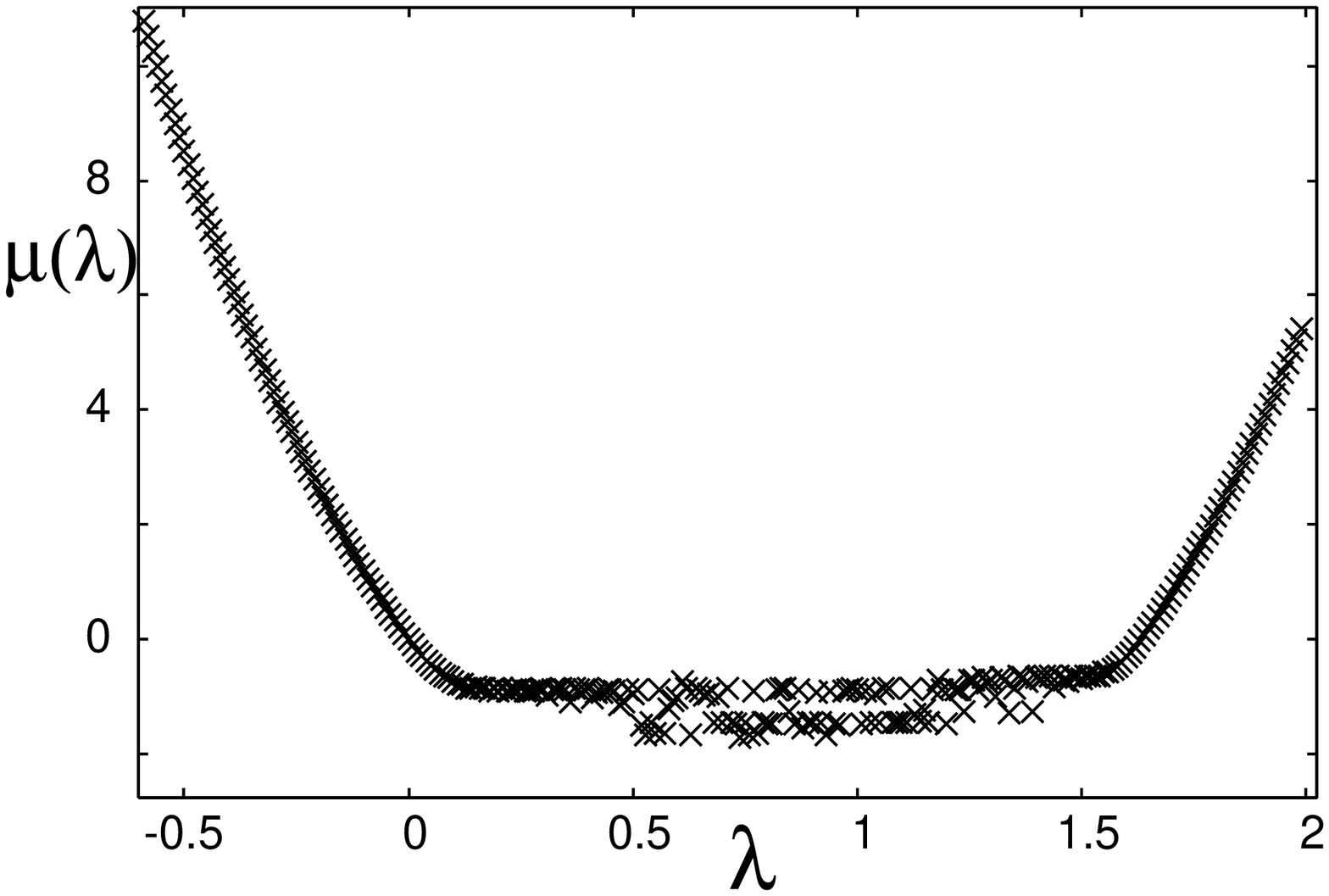}
\caption{$\m(\l)$ when $L=50$, $(\a,\b,\g,\d)=(0.9,0.9,0.1,0.1)$ (maximum-current phase), and $p_l=0.2$.}
\label{fig14}
\end{center}
\end{figure}

\begin{figure}
\begin{center}
\includegraphics[scale=0.4]{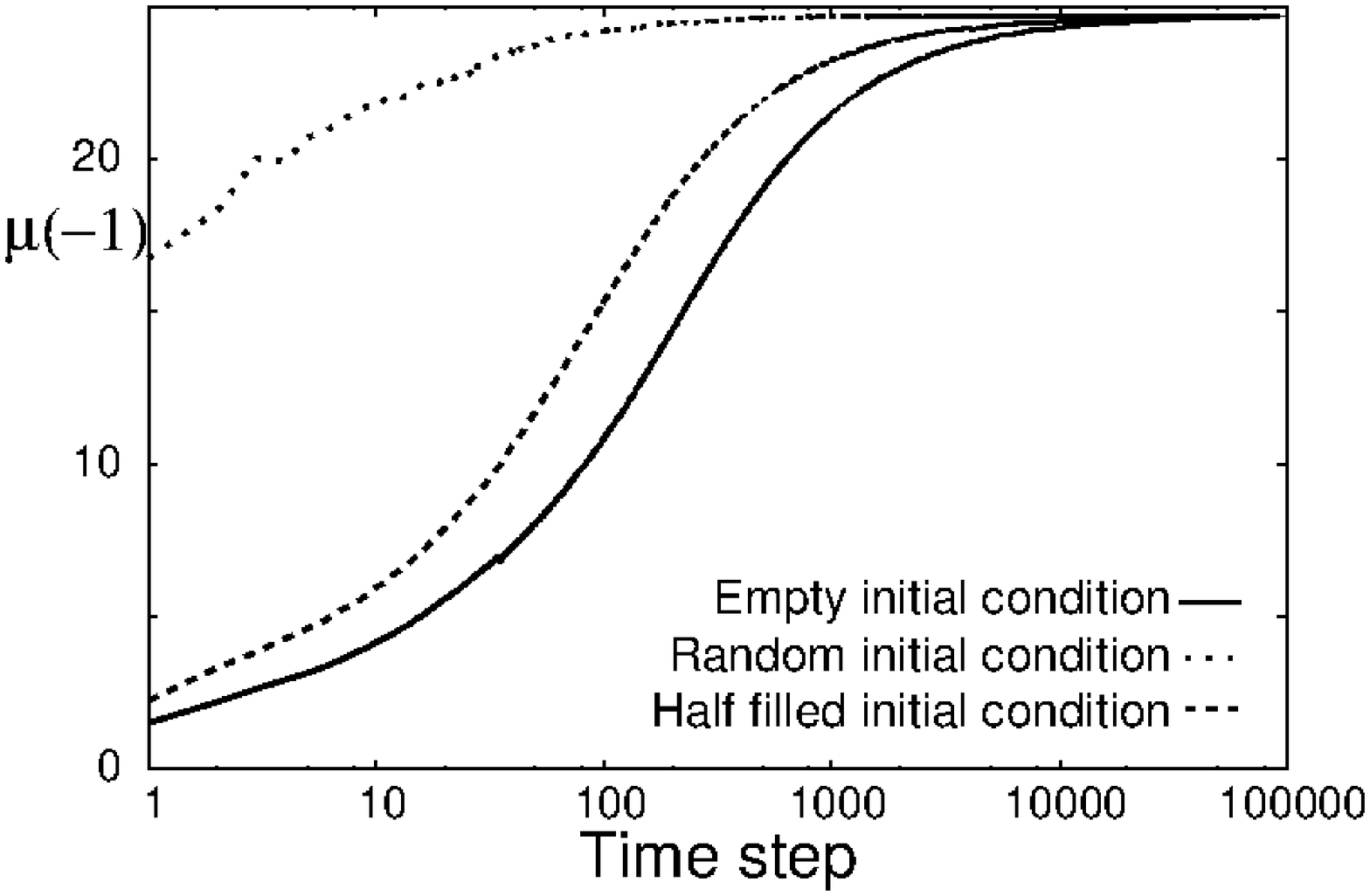}
\includegraphics[scale=0.4]{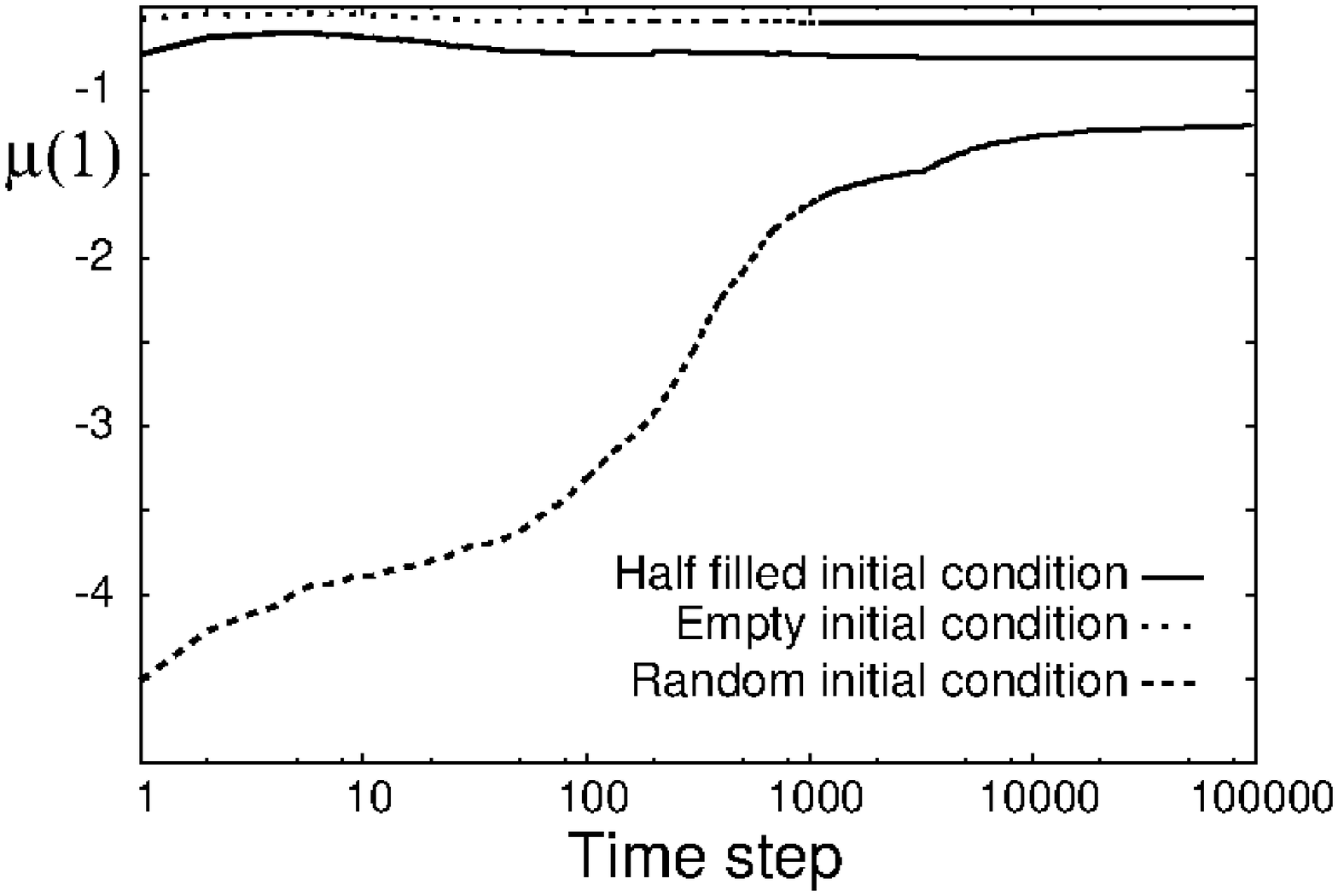}
\caption{Time evolutions of $\m(-1)$ (top) and $\m(1)$ (bottom) when
 $L=50$ and $p_l=0.2$ starting from three initial conditions: empty, random,
 and half-filled.}
\label{fig15}
\end{center}
\end{figure}

The Monte Carlo method
sometimes presents non-negligible errors.  It happens when the hopping asymmetry
and the system size are large.
For example, Fig. \ref{fig14} illustrates the instability observed in the
numerically obtained generating function.  
Here, we have chosen the parameters  
$L=50$, $(\a,\b,\g,\d)=(0.9,0.9,0.1,0.1)$ (maximum-current
phase), and $p_l=0.2$.  This figure shows characteristic
fluctuations in the region $0.2<\l<1.5$.
The true generating function should be a smooth or at least convex
function.
However, the obtained data does not satisfy the expected convexity.
To find the cause of the fluctuations, we investigated the initial condition
dependence of our Monte Carlo simulation. 
In Fig. \ref{fig15}, we show how $t^{-1}\ln(X_1\cdots X_K)$ evolves
with time in the cases of $\l=-1$ and $\l=1$ each with three different initial
configurations, namely, the empty initial condition (all $\t_i=0$), the
random initial condition ($\t_i$ is randomly chosen with probability
1/2), and the half-filled initial condition ($\t_i=1$ for $i<L/2$ and
$\t_i=0$ for the other value of $i$).
The other parameters used are the same as those indicated in Fig. \ref{fig14}.
Note that $\l=-1$ corresponds to the stable region and $\l=1$ the
unstable region.
The top of Fig. \ref{fig15} shows the case $\l=-1$, where 
the three samples
exhibit convergence to a common value
at $t=10^4$.  
The bottom shows the case $\l=1$, where the convergence is very slow and
the initial-condition dependence remains even at $t=10^5$.  
This slow convergence is considered to be related to
the instability shown in Fig. \ref{fig14}.

\begin{figure}
\begin{center}
\includegraphics[scale=0.4]{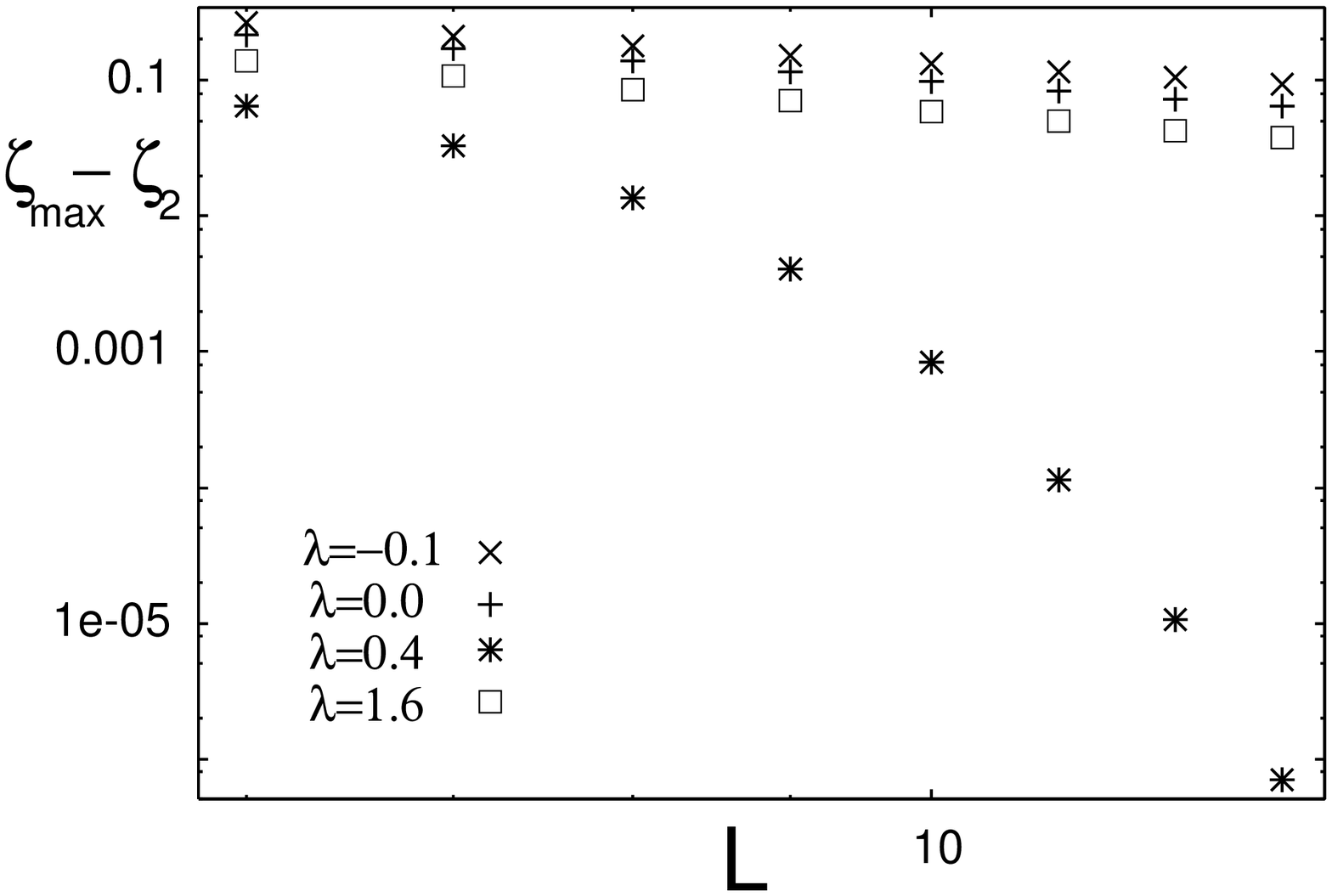}
\includegraphics[scale=0.4]{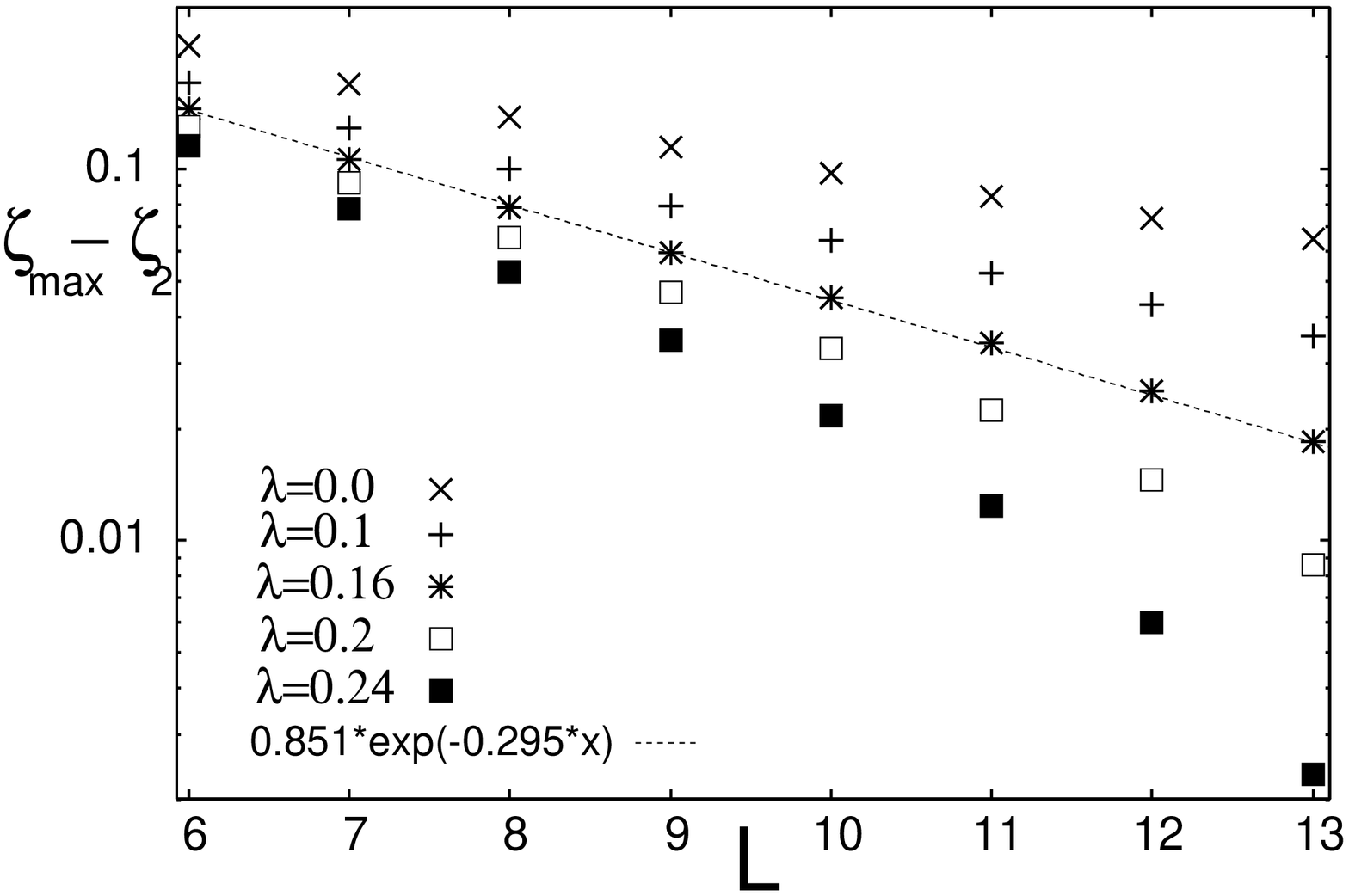}
\caption{$\z_{\mathrm{max}}-\z_2$ versus the system size $L=6,\cdots,13$.  The
 top figure shows the cases $\l=-0.1$, $0.0$, $0.4$, and $1.6$ in the log-log scale
 and the bottom figure shows the cases $\l=0.0$, $0.1$, $0.16$, $0.2$, and $0.24$ in
 the semi-log scale.  The dotted line in the bottom figure is a
 fitted exponential function.}
\label{fig16}
\end{center}
\end{figure}

The convergence property should be reflected in the spectrum of eigenvalues
of the $\l$-modified transition matrix $\mathsf{W}^\l$.  Thus,
we calculated the second largest eigenvalue $\z_2$.
In Fig. \ref{fig16}, we show the system size dependence of the gap
$\z_{\mathrm{max}}-\z_2$ for several $\l$ values in the systems of sizes
$L=6,\cdots,13$, where $\z_{\mathrm{max}}$ is the largest eigenvalue.
The top figure shows the cases $\l=-0.1$, $0.0$, $0.4$, and $1.6$ on the log-log scale
and the bottom figure shows the cases $\l=0.0$, $0.1$, $0.16$, $0.2$, and $0.24$ on
the semi-log scale.
The figure shows that the magnitude of the gap is proportional 
to a certain negative power of $L$ in the stable region, whereas the gap decreases 
faster than the exponential decrease with $L$ in the unstable region. 
In the bottom of Fig. \ref{fig16}, we see that
the system size dependence of $\z_{\mathrm{max}}-\z_2$ changes from the power
to exponential and finally to stretched exponential-like.
Thus, one reason for the instability of the Monte Carlo calculation found here 
is considered to be that
$\z_2$ becomes close to $\z_{\mathrm{max}}$ as $L$ increases.
This behavior suggests that a certain type of phase transition may occur
as reported recently\cite{Rak08,Tch10}.

\section{Discussion and conclusions}

Our simulation results show $\m_A^N(\l)=\m_B^N(\l)$, which implies
\begin{equation}
\bra q_A^n \ket_c\sim L^n \bra q_B^n\ket_c.
\end{equation}
From the calculation given by Bodineau and Derrida for the SSEP\cite{Der04}, we may
assume $\bra q_B^n\ket_c\sim\frac{1}{L}$, and thus 
$\bra q_A^n\ket_c\sim L^{n-1}$.
This relation does not coincide with our result 
$\bra q_A^{2n} \ket_c\sim L$.
These conflicting results suggest the necessity of further study.
The calculation by Bodineau and Derrida assumes the additivity
principle, the scaling relation $f(q_B)\sim g(q_BL)/L$, and the large deviation
locally defined to be quadratic.
Our results do not support the scaling relation and
the Gaussian property of the local large deviation function.
Both assumptions may need revision.
However, our results have some drawbacks.
The estimation by Bodineau and Derrida is given in the entire parameter
region, while our simulation is limited to several sets of parameters.
Furthermore, the present Monte Carlo method has errors in several cases.
The improvement of the Monte Carlo method to obtain the generating
function is necessary.
We must note that the DMRG method is another method for obtaining a large
deviation\cite{Gor09}, and not only the continuous time, which is well
studied, but also the discrete time case based on the DMRG method for the steady
state\cite{Hie98} can be established.

The relation $q_A\sim Lq_B$ can be derived from the plausible argument
considering the time-dependent and time-independent contributions to the
current.
The details are presented in $\S$ 2.
Actually, the two generating functions for the two currents
coincide with each other in relatively small systems, as
shown in Fig. 9.  Although they are considered to be the same
also in large systems, we could not confirm it because the
Monte Carlo calculation is not suitable for determining $q_B$.  Thus, in a
practical sense, they have differences.

We observed the cusp around $q=0$ in the current large deviation 
function in the ASEP.
A similar cusp in the large deviation function for entropy production 
was reported by Mehl et al.\cite{Meh08} 
in a system of an overdamped Brownian particle under the periodic
potential and uniform external field.
They attributed the generation of the cusp to the sublinear response of
the particle to the external field, which was derived from the
potential within the site.
In our study for the ASEP, the sublinear response may be caused by the
interactions of particles, where the cause is different from that in the
case of a one-particle system.
The physical basis of the cusp we consider is that the probability of
observing the current toward a lower hopping rate becomes $0$, which
means that the current becomes zero in that direction.
This explains why the cusp is observed, though this does not explain how
the cusp is generated.

In conclusion, in this study we have calculated numerically the current large deviation 
function in the SSEP and ASEP with open boundaries.
We have found that the generating function and the large deviation function 
deviate from the quadratic form.
The system size dependence of the even order cumulants is found to be 
proportional to $L$ in both the SSEP and the ASEP.
The generating functions defined by different definitions of the current
appear to be the same for small systems.
For the large system size, the Monte Carlo simulation gives unsatisfying
results for the current through a boundary, from which we conjecture
that the use of the total current is suitable for the Monte Carlo
calculation of the large deviation function.
The symmetry relation for the total current is calculated 
both numerically and analytically, and the obtained results are in good agreement.
For the Monte Carlo calculations, a small deviation is observed in the
symmetry relations and the deviation depends on the boundary parameters.
We found that the convergence of the Monte Carlo simulation for
calculating the large deviation function in the ASEP depends on the system
size and the hopping parameter of a particle.
The convergence is not good when the system size is large and the
asymmetry of the hopping is large.
In the region where the convergence is not good, the difference between
the largest and second largest eigenvalues of the modified
transition matrix depends on the system size as the stretched exponential,
which is faster than the exponential decay.

\section*{Acknowledgments}

One of the authors (T.M) wishes to express gratitude to Hisao Hayakawa
for all the support and helpful discussions.
We would like to acknowledge Yasuhiro Hieida for giving us many helpful comments on
the paper.
The numerical calculations were carried out at YITP in Kyoto University.
This work was supported by the Grant-in-Aid for the Global COE Program
"The Next Generation of Physics, Spun from Universality and Emergence"
from the Ministry of Education, Culture, Sports, Science and Technology
(MEXT) of Japan.


\begin{thebibliography}{10,<num>}
\bibitem{referee} The argument here is constructed as per suggestion of the
	anonymous referee.
\end{thebibliography}
\end{document}